\documentclass[a4paper]{article}

\usepackage{amsmath,amsfonts,amssymb,amsthm,latexsym}
\usepackage[textures]{graphics}
\usepackage{epsf}

\ifeof5
    \def\axocolor{ }
    \def\SetColor#1{ }

\else
    \input{colordvi.sty}
    \background{White}
    \textBlack
    \def\axocolor{Black }
    \def\SetColor#1{\def\axocolor{#1 }}
	
\fi
\def\axowidth{0.5 }
\def\axoscale{1.0 }
\def\axoxoff{0 }
\def\axoyoff{0 }
\def\axoxo{0 }
\def\axoyo{0 }
\def\firstcall{1}
\def\Gluon(#1,#2)(#3,#4)#5#6{
\put(\axoxoff,\axoyoff){
}
}
\def\Photon(#1,#2)(#3,#4)#5#6{
\put(\axoxoff,\axoyoff){
}
}
\def\ZigZag(#1,#2)(#3,#4)#5#6{
\put(\axoxoff,\axoyoff){
}
}
\def\PhotonArc(#1,#2)(#3,#4,#5)#6#7{
\put(\axoxoff,\axoyoff){
}
}
\def\GlueArc(#1,#2)(#3,#4,#5)#6#7{
\put(\axoxoff,\axoyoff){
}
}
\def\ArrowArc(#1,#2)(#3,#4,#5){
\put(\axoxoff,\axoyoff){
}
}
\def\LongArrowArc(#1,#2)(#3,#4,#5){
\put(\axoxoff,\axoyoff){
}
}
\def\DashArrowArc(#1,#2)(#3,#4,#5)#6{
\put(\axoxoff,\axoyoff){
}
}
\def\ArrowArcn(#1,#2)(#3,#4,#5){
\put(\axoxoff,\axoyoff){
}
}
\def\LongArrowArcn(#1,#2)(#3,#4,#5){
\put(\axoxoff,\axoyoff){
}
}
\def\DashArrowArcn(#1,#2)(#3,#4,#5)#6{
\put(\axoxoff,\axoyoff){
}
}
\def\ArrowLine(#1,#2)(#3,#4){
\put(\axoxoff,\axoyoff){
}
}
\def\LongArrow(#1,#2)(#3,#4){
\put(\axoxoff,\axoyoff){
}
}
\def\DashArrowLine(#1,#2)(#3,#4)#5{
\put(\axoxoff,\axoyoff){
}
}
\def\Line(#1,#2)(#3,#4){
\put(\axoxoff,\axoyoff){
}
}
\def\DashLine(#1,#2)(#3,#4)#5{
\put(\axoxoff,\axoyoff){
}
}
\def\CArc(#1,#2)(#3,#4,#5){
\put(\axoxoff,\axoyoff){
}
}
\def\DashCArc(#1,#2)(#3,#4,#5)#6{
\put(\axoxoff,\axoyoff){
}
}
\def\Vertex(#1,#2)#3{
\put(\axoxoff,\axoyoff){
}
}
\def\Text(#1,#2)[#3]#4{
\dimen0=\axoxoff \unitlength
\dimen1=\axoyoff \unitlength
\advance\dimen0 by #1 \unitlength
\advance\dimen1 by #2 \unitlength
\makeatletter
\@killglue\raise\dimen1\hbox to\z@{\kern\dimen0 \makebox(0,0)[#3]{#4}\hss}
\ignorespaces
\makeatother
}
\def\BCirc(#1,#2)#3{
\put(\axoxoff,\axoyoff){
}
}
\def\GCirc(#1,#2)#3#4{
\put(\axoxoff,\axoyoff){
}
}
\def\CCirc(#1,#2)#3#4#5{
\put(\axoxoff,\axoyoff){
}
\put(\axoxoff,\axoyoff){
}
}
\def\EBox(#1,#2)(#3,#4){
\put(\axoxoff,\axoyoff){
}
}
\def\BBox(#1,#2)(#3,#4){
\put(\axoxoff,\axoyoff){
}
}
\def\GBox(#1,#2)(#3,#4)#5{
\put(\axoxoff,\axoyoff){
}
}
\def\CBox(#1,#2)(#3,#4)#5#6{
\put(\axoxoff,\axoyoff){
}
\put(\axoxoff,\axoyoff){
}
}
\def\Boxc(#1,#2)(#3,#4){
\put(\axoxoff,\axoyoff){
}
}
\def\BBoxc(#1,#2)(#3,#4){
\put(\axoxoff,\axoyoff){
}
}
\def\GBoxc(#1,#2)(#3,#4)#5{
\put(\axoxoff,\axoyoff){
}
}
\def\CBoxc(#1,#2)(#3,#4)#5#6{
\put(\axoxoff,\axoyoff){
}
\put(\axoxoff,\axoyoff){
}
}
\def\ETri(#1,#2)(#3,#4)(#5,#6){
\put(\axoxoff,\axoyoff){
}
}
\def\BTri(#1,#2)(#3,#4)(#5,#6){
\put(\axoxoff,\axoyoff){
}
}
\def\GTri(#1,#2)(#3,#4)(#5,#6)#7{
\put(\axoxoff,\axoyoff){
}
}
\def\CTri(#1,#2)(#3,#4)(#5,#6)#7#8{
\put(\axoxoff,\axoyoff){
}
\put(\axoxoff,\axoyoff){
}
}
\def\SetWidth#1{\def\axowidth{#1 }}

\def\SetOffset(#1,#2){\def\axoxoff{#1 } \def\axoyoff{#2 }}
\def\SetScaledOffset(#1,#2){\def\axoxo{#1 } \def\axoyo{#2 }}
\def\pfont{Times-Roman }
\def\fsize{10 }
\makeatletter
\def\fmode{4 }
\def\@l@{l} \def\@r@{r} \def\@t@{t} \def\@b@{b}
\def\mymodetest#1{\ifx#1\end \let\next=\relax \else {
\if#1\@r@\global\def\fmodeh{1 }\fi
\if#1\@l@\global\def\fmodeh{-1 }\fi
\if#1\@b@\global\def\fmodev{3 }\fi
\if#1\@t@\global\def\fmodev{-3 }\fi
} \let\next=\mymodetest\fi \next}
\makeatother
\def\PText(#1,#2)(#3)[#4]#5{
\def\fmodev{0 }
\def\fmodeh{0 }
\mymodetest#4\end
\put(\axoxoff,\axoyoff){\makebox(0,0)[]{\special{"/\pfont findfont \fsize
 scalefont setfont \axocolor #1 \axoxo add #2 \axoyo add #3
\fmode \fmodev add \fmodeh add \fsize (#5) \axoscale ptext }}}
}
\def\GOval(#1,#2)(#3,#4)(#5)#6{
\put(\axoxoff,\axoyoff){
}
}
\def\COval(#1,#2)(#3,#4)(#5)#6#7{
\put(\axoxoff,\axoyoff){
}
\put(\axoxoff,\axoyoff){
}
}
\def\Oval(#1,#2)(#3,#4)(#5){
\put(\axoxoff,\axoyoff){
}
}
\let\eind=]

\def\kromme(#1,#2)#3{#1 \axoxo add #2 \axoyo add \ifx #3\eind\else
\expandafter\kromme\fi#3}
\def\LogAxis(#1,#2)(#3,#4)(#5,#6,#7,#8){
\put(\axoxoff,\axoyoff){
}
}
\def\LinAxis(#1,#2)(#3,#4)(#5,#6,#7,#8,#9){
\put(\axoxoff,\axoyoff){
}
}
\input rotate.tex
\makeatletter
\def\rText(#1,#2)[#3][#4]#5{
\ifnum\firstcall=1\global\def\firstcall{0}\rText(-10000,#2)[#3][]{#5}\fi
\dimen2=\axoxoff \unitlength
\dimen3=\axoyoff \unitlength
\advance\dimen2 by #1 \unitlength
\advance\dimen3 by #2 \unitlength
\@killglue\raise\dimen3\hbox to \z@{\kern\dimen2
\makebox(0,0)[#3]{
\ifx#4l{\setbox3=\hbox{#5}\rotl{3}}\else{
\ifx#4r{\setbox3=\hbox{#5}\rotr{3}}\else{
\ifx#4u{\setbox3=\hbox{#5}\rotu{3}}\else{#5}\fi}\fi}\fi}\hss}
\ignorespaces
}
\makeatother
\def\BText(#1,#2)#3{
\put(\axoxoff,\axoyoff){
}
}
\def\GText(#1,#2)#3#4{
\put(\axoxoff,\axoyoff){
}
}
\def\CText(#1,#2)#3#4#5{
\put(\axoxoff,\axoyoff){
}
\put(\axoxoff,\axoyoff){
}
}
\def\B2Text(#1,#2)#3#4{
\put(\axoxoff,\axoyoff){
}
}
\def\G2Text(#1,#2)#3#4#5{
\put(\axoxoff,\axoyoff){
}
}
\def\C2Text(#1,#2)#3#4#5#6{
\put(\axoxoff,\axoyoff){
}
\put(\axoxoff,\axoyoff){
}
}

\usepackage{axodraw}

\usepackage{color}

\def\br{\begin{eqnarray}}
\def\er{\end{eqnarray}}
\def\brs{\begin{eqnarray*}}
\def\ers{\end{eqnarray*}}
\def\be{\begin{equation}}
\def\ene{\end{equation}}
\def\bd{\begin{description}}
\def\ed{\end{description}}

\def\pa{\partial}

\def\a{\alpha}
\def\b{\beta}
\def\d{\delta}
\def\D{\Delta}
\def\eps{\varepsilon}
\def\g{\gamma}
\def\G{\Gamma}

\def\l{\lambda}

\def\m{\mu}
\def\n{\nu}

\def\P{\Phi}
\def\pa{\partial}

\def\s{\sigma}
\def\S{\Sigma}

\def\th{\theta}

\def\sqr#1#2{{\vcenter{\vbox{\hrule height.#2pt
         \hbox{\vrule width.#2pt height#1pt \kern#1pt
            \vrule width.#2pt}
         \hrule height.#2pt}}}}

\makeatletter
\def\secteqno{\@addtoreset{equation}{section}%
\def\theequation{\thesection.\arabic{equation}}}

\begin{document}
\secteqno

\renewcommand{\thefootnote}{\fnsymbol{footnote}}

\noindent

\begin{titlepage}

\vspace*{20pt}
\bigskip

\begin{center}
{ \Large \bf Two loop divergences studied\\ with one loop Constrained Differential Renormalization   }
\end{center}

\bigskip

\vskip 0.8truecm

\centerline{\sc
Cesar Seijas\footnote{\tt cesar@fpaxp1.usc.es}}

\vspace{1pc}

\begin{center}
{\em
Departamento de F\'\i sica de Part\'\i culas, Universidade de
Santiago de Compostela,\\
E-15706 Santiago de Compostela, Spain\\}

\vspace{5pc}

{\large \bf Abstract}

\end{center}

\noindent
In the context of Differential Renormalization, using Constrained Differential Renormalization rules at one loop, we show how to obtain 
concrete results in two loop calculations without making use of Ward identities. In order to do that, we obtain a list of integrals
with overlapping divergences compatible with CDR that can be applied to various two loop background field calculations. As an
example, we obtain the two loop coefficient of the beta function of QED, SuperQED and Yang-Mills theory.

\end{titlepage}

\renewcommand{\thefootnote}{\arabic{footnote}}
\setcounter{footnote}{0}


\section{Introduction}

Since its introduction in 1991  \cite{Freedman:1991tk}  differential renormalization (DR) has 
been applied in a large
variety of situations like $\phi^4$ at three loops \cite{Freedman:1991tk}, 
Yang Mills at one loop \cite{Freedman:1991tk}, QED at two loops
\cite{Haagensen:1992vz}, the Wess-Zumino model at three loops \cite{Haagensen:1991vd}, 
SuperQED at two loops \cite{Song} and SuperYM at
two loops \cite{Mas:2002xh}    among others. In all of them the  method has exhibited one
of its strengths: the simple and elegant way of obtaining the coefficients of the 
renormalization group equation (beta functions
and anomalous dimensions). 

The classical objection to the method points out  that  the renormalization 
of each divergence introduces an ``a
priori" independent scale. In order to preserve gauge invariance one has 
to apply at  each order Ward identities in
order to obtain the correct relations among scales. A constructive set of 
rules that avoids this somewhat tedious
procedure has only been deviced at one loop. The enhanced set of rules that 
goes by the name of Constrained
Differential Renormalization (CDR)
\cite{delAguila:1997kw,delAguila:1997su,delAguila:1998nd,Perez-Victoria:1998fj}  
automatically produces
expressions with a single mass scale that fulfill Ward identities. 

At two loops a fully consistent extension of CDR is not known. However we believe that the results of this paper are a small
step ahead in the right direction. Mainly, we will be dealing with the divergent parts of the amplitudes. For them
we will show how one-loop CDR still has the strength to produce gauge invariant answers. Interestingly enough, from the divergent parts
useful results can still be obtained such as the beta function and the anomalous dimensions.

The structure of the paper is as follows. First we  briefly review the basic rules of  DR and   CDR. In the next section we write
a list of renormalized expressions for several two-loop integrals that contain overlapping divergences and comment about the general
strategy to obtain them. As an example of the use of these integrals, we obtain the two loop coefficient of the beta function
of QED, SuperQED and Yang-Mills. The first two calculations basically reproduce the results obtained in \cite{Haagensen:1992vz}
   and \cite{Haagensen:1991vd} in a more efficient way. 

\section{DR and CDR}
\subsection{Differential Renormalization (DR)}

DR is a renormalization method that consists in replacing   coordinate-space expressions that are too singular by derivatives of
less singular ones. This method does not need cutoff nor explicit counterterms, although they are implicitly used when performing formal
integration by parts.
The basic idea is that divergent one loop expressions are well defined for non-coincident points, but at short 
distances the amplitude is too singular and does not have a Fourier transform. So, to renormalize one can replace the divergent
expression for the derivative of a less singular one that has the same values as the original outside the origin, and has a well
defined Fourier transform (if formal integration by parts is used with the derivatives).

This method is especially well suited for supersymmetric theories 
because we stay in four dimensions all the time, which is not the case with other methods as dimensional regularization or reduction.

As an example consider the one loop contribution of $\phi^4$ theory. The bare expression is
\br
\G (x_1 , x_2 , x_3, x_4 ) &=& \frac{\l^2}{32 \pi^4} \left[ \d (x_1 - x_2 )\d (x_3 - x_4) \frac{1}{(x_1-x_4)^4} + (2\; perms) \right]
\er
At short distance $\frac{1}{x^4}$ does not have a well defined Fourier transform, and DR proposes to 
substitute this for
\br
\frac{1}{x^4} \rightarrow \left[\frac{1}{x^4} \right]_R &=& - \frac{1}{4} \Box \frac{ \ln x^2 M^2}{x^2}
\er
Both expressions coincide for $x\neq 0$, but the new one has a well defined Fourier transform if we neglect the
divergent surface terms that appears upon integrating by parts the d'Alembertian. Here is where the
counterterms hide, and by applying formal integration by parts \cite{Freedman:1991tk} we are implicitly taking them into account.
I.e.
\br
\int d^4 x \; e^{i p \cdot x} \left[\frac{1}{x^4} \right]_R &=& - \frac{1}{4} 
\int d^4 x e^{i p \cdot x} \Box \frac{ \ln x^2 M^2}{x^2} = \frac{p^2}{4}
 \int d^4 e^{i p \cdot x} \frac{\ln x^2 M^2}{x^2}
\nonumber \\ &=& - \pi^2 \ln \left( \frac{p^2}{{\bar{M}}^2}\right) 
\er
A constant with mass dimension $M$ has been introduced for dimensional reasons. It parametrizes the {\em local
ambiguity}
$$
 \Box \frac{\ln x^2 {M'}^2}{x^2} =  \Box \frac{\ln x^2 M^2}{x^2}  + 2\ln\frac{M'}{M} ~\delta(x)
 $$
A crutial observation is that this shift $M\to M'$  can be absorbed in a rescaling of the coupling
constant
$\l$
\cite{Freedman:1991tk}. This is a hint that renormalized amplitudes satisfy renormalization group equations, with M
playing the role of the renormalization group scale.

Although we will not treat massive theories here, they have been considered in \cite{Freedman:1991tk} 
and  \cite{Haagensen:1992am}. The appearance of a bare mass does not interfere with the method since the DR is related to
short-distance singularities and masses only change the long-distance behaviour of the correlators.

As was previously mentioned, each divergency is cured independently by a suitable replacement like the one above. An important issue
arises with symmetric theories, where gauge invariance should be satisfied at each order in perturbation theory. The standard lore
is that Ward  identities (except for anomalies) can always be satisfied with the renormalized expression by adjusting the different
mass scales obtained for different diagrams and expressions (ie. fixing a renormalization scheme); but at the same time, one always has to explicitly apply these identities to obtain the correct scale relation.

\subsection{Constrained Differential Renormalization (CDR)}

The basic idea is to obtain a renormalization scheme that automatically preserves Ward identities for renormalized amplitudes.
For this,  the following rules will be applied
\begin{enumerate}
\item {\em Differential reduction}
\begin{itemize}
\item Functions with singular behaviour worse than logarithmic are reduced to derivatives of (at most) logarithmically divergent
functions without introducing extra dimensionful constants.
\item Logarithmically divergent expressions are written as derivatives of regular functions, 
introducing one single constant $M$, which has dimensions of mass and plays the role of the renormalization group scale.
\end{itemize}
\item {\em Formal integration by parts} We do not take care of the surface terms that appear when applying integration by parts. Related to this, differentiation and renormalization must be two commutative operations: let $F$ an arbitrary function, then 
$[ \pa F ]_R = \pa [F]_R$.
\item {\em Renormalization rule of the delta function}:
\begin{equation} [ F (x, x_1, \ldots , x_n ) \d (x-y) ]_R = [ F ( x, x_1, \ldots , x_n)]_R \d (x-y)
\end{equation}
\item Validity of the propagator equation
\begin{equation} [F(x,x_1,\ldots,x_n) ( \Box - m^2) \D_{m}(x)]_R = - [F(x,x_1,\ldots,x_n) \d(x)]_R
\end{equation} 
where $\D_{m}$ is the massive propagator of a particle of mass m and $F$ an arbitrary function.

\end{enumerate}

Central to the fulfillment of the Ward identities (and the action principle, from which they can be derived)
 is that the application of the kinetic differential operator to some propagator line inside a Feynman graph is
equivalent to the contraction of the line to a point. This statement is guaranteed to hold through the previous set of rules.
The upshot is a basic set of renormalized expressions (basic functions) with different numbers of propagators and various
differential operators acting only on one of the propagators, and involving a single scale $M$. Therefore the CDR
program amounts to the following two step operation:
\begin{itemize}
\item
Express the Feynman diagram in terms of these basic functions performing all the index contractions 
 (this is an important point, because CDR does not commute with index contraction) and by means of 
the Leibniz rule moving all the
derivatives to act on one of the propagators.
\item
Replace the basic functions with their renormalized version.
\end{itemize}
Applying the previously defined rules, one can obtain a set of renormalized basic functions. 
Here we present a list of the most relevant ones ($\D$ stands for the massless propagator $\D = \frac{1}{4 \pi^2}\frac{1}{x^2}$).
\br
\D^2_R &=& - \frac{1}{4 (4 \pi^2)^2} \Box \frac{\ln x^2 M^2}{x^2} \nonumber \\
(\D \pa_{\m} \D)_R &=& \frac{1}{2}\pa_{\m} \D^2_R  \nonumber \\
(\D \pa_{\m} \pa_{\n} \D)_R &=& \frac{1}{3} (\pa_{\m} \pa_{\n} - \frac{1}{4} \d_{\m \n} \Box) \D^2_R + \frac{1}{288 
\pi^2} (\pa_{\m} \pa_{\n} - \d_{\m \n} \Box ) \d (x) \nonumber \\
(\D \Box \D)_R &=& 0 \;. 
\er

Out from the list of CDR results for three propagators \cite{{delAguila:1997kw},{delAguila:1998nd}}, in our calculations we will make use of the decomposition into trace and traceless part that is imposed by CDR as
\br
T_R[\pa_{\m} \pa_{\n}] &=& T_R[\pa_{\m} \pa_{\n} - \frac{1}{4} \d_{\m \n} \Box] + \frac{1}{4} \d_{\m \n} T_R [\Box] - 
\frac{1}{128 \pi^2} \d_{\m \n} \d (x) \d (y) \label{CDR_T}
\er

where we have defined $T[{\cal{O}}] = \D \D {\cal{O}} \D $.

When using other gauges different from Feynman gauge, some bare expressions are written in terms of a quantity we define as $\bar{\D} (x) = \frac{1}{4 (4 \pi^2)} \ln x^2 s^2$, where $s$ is an irrelevant constant with mass dimension. For this structure, CDR prescribes \cite{delAguila:1997kw}
\br
( \D \Box \bar{\D} )_R &=& \D^2_R \nonumber \\
( \D \pa_{\m} \pa_{\n} \bar{\D} )_R &=& \frac{1}{4} \left( \D^2_R - \frac{1}{32 \pi^2} \pa_{\m} \pa_{\n} \frac{1}{x^2} \right) \label{CDR_gen_gauge}
\er 

CDR has been checked in abelian and non-abelian gauge theories \cite{delAguila:1997kw,Perez-Victoria:1998fj}, and in supersymmetric calculations \cite{delAguila:1997ma,delAguila:1997yd}.

\section{Two loop use of one loop CDR results}
\label{2loop_CDR}

After CDR has been applied, all the ambiguities (local terms)  at
one loop are fixed in such a way that Ward identities are automatically fulfilled; we will now see how well we can do
in  two loop calculations. We anticipate that at two loops we will be only interested in the divergent part.

Before we proceed with all of this, let us set up some notation that we use in this paper. In our calculations we have found that most of the results can be put in terms of an integral expression that we designate as $I^1$
\br
I^1 (x-y) = \int d^4 u \D_{xu} \D^2_{yu}
\er

which is renormalized applying CDR rules as 

\begin{equation}
I^{1}_R (x) = \frac{1}{4 (4 \pi^2)^2} \frac{\ln x^2 M^2}{x^2} 
\end{equation}

\subsection{Nested divergences}
\label{Nested_div}

This case is particularly simple because CDR can be applied in a systematic way. Starting from the
``inner'' divergency, its regularization according to CDR gives an expression with logarithms of a single scale ($ \ln x^2 M^2 $) and
fixed local terms. The one-loop Ward identities are fulfilled. In the next step, when tackling the outer part of the
diagram, a simple logarithm like the one shown above is promoted to an expression of the form  $ \ln^2 x^2 M^2 + m
\ln x^2 M^2$, with $m$ a calculable coefficient; at the same time, the finite terms multiply outer divergences
which will produce additional logarithms of new scales (the genuine two-loop scales). CDR does note yet prescribe what
the different scales of this later step should be. Hence, we may take all of them the same, and equal to $M$, at the
price of leaving undetermined  finite terms.  In summary, as long as
we are interested in two-loop quantities  that just depend on the logarithms, we see that CDR at one loop 
prescribes a
unique answer. If we were interested in three-loop calculations, the  two loop finite parts would be essential.

This simple scheme has some subtleties when studying diagrams with indices, because index 
contraction does not commute with CDR. Therefore, the correct order is to  insert into the outer
diagram the non-renormalized expression for the ``inner'' one loop diagram, perform all the index contractions, and
then renormalize.

With this procedure we have renormalized all the different structures made up with $I^1$ that we have found in our calculations. The divergent parts of those structures are  (with $\ldots$ standing for the finite local contribution that we are not taking into account)
\br
(\D I^1)_R (x) &=& - \frac{1}{32 (4 \pi^2)^3} \Box \frac{\ln^2 x^2 M^2 + 2 \ln x^2 M^2}{x^2} + \ldots \nonumber \\
(\D \pa_{\m} I^{1})_R (x) &=& - \frac{1}{64 (4 \pi^2)^3} \pa_{\m} \Box \frac{ \ln^2 x^2 M^2 + \ln x^2 M^2}{x^2} + \ldots\nonumber \\
(\D \pa_{\m} \pa_{\n} I^1)_R (x) &=& - \frac{1}{96 (4 \pi^2)^3} \left[\pa_{\m} \pa_{\n} \Box \frac{ \ln^2 x^2 M^2 + \frac{2}{3} \ln x^2 M^2}{x^2} - \frac{1}{4}\d_{\m \n} \Box \Box \frac{\ln^2 x^2 M^2 + \frac{11}{3} \ln x^2 M^2}{x^2} \right] + \ldots \nonumber \\
( \D \Box I^1 )_R (x) &=& \frac{1}{32 ( 4 \pi^2)^2} \Box \Box \frac{ \ln x^2 M^2}{x^2} + \ldots
 \label{2loop_DR_id}
\er

\subsection{Overlapping divergences}
\label{Overlapping_div}

In this section we will focus on the subtle point of the application of CDR results to a two loop calculation: diagrams with overlapping divergences. In these cases it is sometimes difficult to recognize the one loop subdivergences that need to be treated with CDR to start with. The idea will be to obtain through different methods a list of renormalized two loop integrals with overlapping divergences where in each calculation one loop CDR rules have been maintained in every step.

The list is restricted to integrals with at most four derivatives acting on the propagators and two free indices. This is a basis with which we can express all  the overlapping two loop contribution to two point functions in most  theories with derivative couplings in a background field approach (as Yang-Mills, QED, SuperQED or SYM). 

We use the conventions of $z = x-y$ and  $\pa_{\m} \equiv \pa_{\m}^x$. We also define $H$ as
\br
H[{\cal{O}}_1,{\cal{O}}_2 \; ; \; {\cal{O}}_3,{\cal{O}}_4] = \int d^4 u d^4 v \; ( {\cal{O}}_1^{x} \D_{xu})( {\cal{O}}_2^{x} \D_{xv})( {\cal{O}}_3^{y} \D_{yu} ) ({\cal{O}}_4^{y} \D_{yv}) \D_{uv}
\er

being ${\cal{O}}_i$ a differential operator. Finally, let us remark that as with nested divergences, we will not obtain the finite parts of these expressions (we are only interested in the divergent parts, which are unambiguously fixed by one loop CDR). In the final results $\ldots$ will stand for these terms.

\br
H^R[1,1 \; ; \; 1,1] &=& \frac{6 \pi^4 \xi(3) }{ ( 4 \pi^2)^4} \D  \equiv a \D \label{int1} \\
H^R[\pa_{\m},1 \; ; \; 1,1] &=&  \frac{ 3 \xi(3)}{16 (4 \pi^2)^2} ( \pa_{\m} \D) \equiv \frac{a}{2} \pa_{\m} \D \label{int2} \\
H^R[1,\pa_{\l} \; ; \; 1,\pa_{\l}] &=&  - \frac{1}{16(4 \pi^2)^3} \Box \frac{\ln z^2 M^2}{z^2} + \ldots \label{int3} \\
\pa_{\l}^x H^R[1,\pa_{\m} \; ; \; 1,\pa_{\l}] &=&  - \frac{1}{32 (4 \pi^2)^3} \pa_{\m} \Box \frac{ \frac{1}{2} \ln z^2 M^2}{z^2} + \dots \label{int4}\\
\pa_{\l}^x H^R[1,1 \; ; \; \pa_{\l} \pa_{\n},1] &=& \frac {1}{32(4 \pi^2)^3} \pa_{\n} \Box \frac{ \frac{1}{4} \ln^2 z^2 M^2 + \frac{3}{4} \ln z^2 M^2 }{z^2} + \ldots \label{int5}\\ 
H^R[1,\pa_{\l} \; ; \; \pa_{\l} \pa_{\m},1] &=&  \frac{1}{32 (4 \pi^2)^3} \pa_{\m} \Box \frac{\frac{1}{8} \ln^2 z^2 M^2 - \frac{7}{8} \ln z^2 M^2 }{z^2} + \ldots \label{int6} \\
H^R[\pa_{\m} \pa_{\l},\pa_{\l} \; ; \; 1,1] &=&  \frac{1}{32(4 \pi^2)^3} \pa_{\m} \Box \frac{ - \frac{1}{2} \ln^2 z^2 M^2 - \ln z^2 M^2}{z^2}+ \ldots  \label{int7} \\
\pa_{\l}^x H^R[1,\pa_{\m} \; ; \; \pa_{\n} \pa_{\l},1] &=& \frac{1}{32(4 \pi^2)^3} \left[ \pa_{\m} \pa_{\n} \Box \frac{ \frac{1}{8} \ln^2 z^2 M^2 + \frac{1}{8} \ln z^2 M^2}{z^2} + \d_{\m \n} \Box \Box \frac{-\frac{1}{4} \ln z^2 M^2}{z^2} \right] + \ldots \nonumber \\ \label{int8}\\
H^R[1,\pa_{\m} \; ; \; 1,\pa_{\n}] &=& \frac{1}{32 (4 \pi^2)^3} \d_{\m \n} \Box \frac{- \frac{1}{2} \ln z^2 M^2}{z^2} + \ldots \label{int9} \\ 
\pa_{\l}^x H^R[1,\pa_{\l} \; ; \; \pa_{\m} \pa_{\n},1] &=& \frac{1}{32 (4 \pi^2)^3} \left[ \pa_{\m} \pa_{\n} \Box \frac{ - \frac{1}{2} \ln z^2 M^2}{z^2} + \d_{\m \n} \Box \Box \frac{\frac{1}{8} \ln^2 z^2 M^2 + \frac{3}{8} \ln z^2 M^2}{z^2} \right] \nonumber + \ldots \\ \label{int10}\\
\pa_{\l}^x H^R[1,\pa_{\l} \; ; \; 1, \pa_{\m} \pa_{\n}] &=& \frac{1}{32 (4 \pi^2)^3} \left[ \pa_{\m} \pa_{\n} \Box \frac{ \frac{1}{2} \ln z^2 M^2}{z^2} + \d_{\m \n} \Box \Box \frac{\frac{1}{8} \ln^2 z^2 M^2 + \frac{3}{8} \ln z^2 M^2}{z^2} \right] \nonumber + \ldots \\ \label{int10a}\\
H^R[1,1 \; ; \; \pa_{\m} \pa_{\n},1] &=&  \frac{1}{32(4 \pi^2)^3} \d_{\m \n} \Box \frac{ \frac{1}{4} \ln^2 z^2 M^2 + \frac{3}{4} \ln z^2 M^2}{z^2} + \ldots \label{int11} \\
\pa_{\l}^x H^R[1,1 \; ; \; \pa_{\l} \pa_{\n},\pa_{\m}] &=& \frac{1}{32 (4 \pi^2)^3} \d_{\m \n} \Box \Box \frac{ \frac{1}{8} \ln^2 z^2 M^2 + \frac{3}{8} \ln z^2 M^2}{z^2} + \ldots \label{int12} \\
\pa_{\l}^x H^R[1,1 \; ; \; \pa_{\m} \pa_{\n},\pa_{\l}] &=& \frac{1}{32 (4 \pi^2)^3} \pa_{\m} \pa_{\n} \Box \frac{\frac{1}{8} \ln^2 z^2 M^2 + \frac{3}{8} \ln z^2 M^2}{z^2} + \ldots \label{int13} \\
H^R[1,\pa_{\m} \pa_{\l} \; ; \; \pa_{\n} \pa_{\l},1] &=& \frac{1}{32 (4 \pi^2)^3} \left[ \pa_{\m} \pa_{\n} \Box \frac{ \frac{1}{6} \ln^2 z^2 M^2 - \frac{5}{36} \ln z^2 M^2}{z^2} + \right. \nonumber \\
&+& \left. \d_{\m \n} \Box \Box \frac{ - \frac{1}{24} \ln^2 z^2 M^2 - \frac{29}{72} \ln z^2 M^2}{z^2} \right] + \ldots \label{int14}\\
H^R[1,\pa_{\m} \pa_{\l} \; ; \; 1,\pa_{\n} \pa_{\l}] &=&  \frac{1}{32 (4 \pi^2)^3} \left[ \pa_{\m} \pa_{\n} \Box \frac{\frac{1}{6} \ln^2 z^2 M^2 + \frac{49}{36} \ln z^2 M^2}{z^2} + \right. \nonumber \\
&+& \left.  \d_{\m \n} \Box \Box \frac{- \frac{1}{24} \ln^2 z^2 M^2 - \frac{11}{72} \ln z^2 M^2}{z^2} \right] + \ldots \nonumber \\ \label{int15}
\er

These integrals are obtained basically by applying two properties:

\begin{itemize}
\item Integral relations presented in appendix \ref{ap_int_rel}. These exact relations allow us to put some of the integrals in terms of others that have an explicit d'Alembertian acting on one of the propagators. Once we have done that, we can apply $\Box \D = - \d$ to put these integrals in terms of the previously defined $I^1$. Then, we can straightforwardly apply the procedure for renormalization with nested divergences that preserves one loop CDR.\footnote{Also this is the reason why we have not listed here the cases where the differential operator is a d'Alembertian. For example, it is obvious that $ H[ \Box,1 \; ; \; 1,1] = - \D I^1$. }

\item The decomposition into trace part, traceless part and fixed local term imposed by CDR to $T[\pa_{\m} \pa_{\n}]$ as (\ref{CDR_T}).

\end{itemize}

Let us show in an explicit example how we apply these two procedures. Considering integral (\ref{int5}), this can be evaluated with both methods. First, we will make use of (\ref{rel_int2}) and put this integral as sum of different integrals that have the divergences nested 
\br
\pa_{\l}^x \int &d^4 u d^4 v & \D_{xu} \D_{xv} ( \pa_{\l}^y \pa_{\n}^y \D_{yu} ) \D_{yv} \D_{uv} = \nonumber \\
&=& - \frac{1}{2} \pa_{\n}^y \int d^4 u d^4 v \; \D_{xu} \D_{xv} ( \Box \D_{yu} ) \D_{yv} \D_{uv} + \nonumber \\
& & + \int d^4 u d^4 v \; \D_{xu} \D_{xv} ( \Box \D_{yu} ) ( \pa_{\n}^y \D_{yv}) \D_{uv} - \nonumber \\
& & - \frac{1}{2} \pa_{\n}^y \pa_{\l}^y \int d^4 u d^4 v \; \D_{xu} \D_{xv} ( \pa_{\l}^y \D_{yu} ) \D_{yv} \D_{uv} \nonumber \\
\er

Now, we have to apply $\Box \D = - \d $ and rewrite this integrals in terms of $I^1$ (note that $a=\frac{6 \pi^4 \xi(3) }{ ( 4 \pi^2)^4}$). The third integral can be easily shown to be finite, and its value is obtained in appendix \ref{ap_integrales}.
\br
\pa_{\l}^x \int &d^4 u d^4 v & \D_{xu} \D_{xv} ( \pa_{\l}^y \pa_{\n}^y \D_{yu} ) \D_{yv} \D_{uv} = \nonumber \\
&=& - \frac{1}{2} \pa_{\n} ( \D I^1 ) + \frac{1}{2} ( \D \pa_{\n} I^1 ) + \frac{a}{4} \pa_{\n} ( \Box \D) \nonumber \\
\er

Applying the renormalization procedure for nested divergences, the final renormalized expression is found to be
\br
\pa_{\l}^x \int &d^4 u d^4 v & \D_{xu} \D_{xv} ( \pa_{\l}^y \pa_{\n}^y \D_{yu} ) \D_{yv} \D_{uv} = \nonumber \\
&=& \frac{1}{32 (4 \pi^2)^3} \pa_{\n} \Box \frac{ \frac{1}{4} \ln^2 z^2 M^2 + \frac{3}{4} \ln z^2 M^2}{z^2} + \ldots
\er 

Where $\ldots$ stand again for the finite terms that we are not taking into account and $z=x-y$. 

The other method of obtaining this integral is making use of the CDR relation (\ref{CDR_T}) and perform a trace-traceless decomposition of $( \pa_{\l}^y \pa_{\n}^y \D_{yu} ) \D_{yv} \D_{uv}$ adding up the CDR term, i.e.
\br
\pa_{\l}^x \int &d^4 u d^4 v & \D_{xu} \D_{xv} ( \pa_{\l}^y \pa_{\n}^y \D_{yu} ) \D_{yv} \D_{uv} = \nonumber \\
&\stackrel{R}{\rightarrow}&  \left. \frac{1}{4} \pa_{\n}^x \int d^4 u d^4 v \; \D_{xu} \D_{xv} ( \Box \D_{yu}) \D_{yv} \D_{uv} \right|_R + \nonumber \\
& & + \left. \pa_{\l}^x \int d^4 u d^4 v \; \D_{xu} \D_{xv} \left[ ( \pa_{\l}^y \pa_{\n}^y - \frac{1}{4} \d_{\l \n} \Box ) \D_{yu} \right] \D_{yv} \D_{uv} \right|_R - \nonumber \\
& & - \frac{1}{128 \pi^2} \pa_{\n}^x \int d^4 u d^4 v \; \D_{xu} \D_{xv} \d (y-u) \d (y-v) \nonumber \\
&=& - \frac{1}{4} \pa_{\n} ( \D I^1 )_R - \frac{1}{128 \pi^2} \pa_{\n} \D^2_R + \pa_{\l} I_{\l \n \; R} \\
&=& \frac{1}{32 (4 \pi^2)^3} \pa_{\n} \Box \frac{ \frac{1}{4} \ln^2 z^2 M^2 + \frac{3}{4} \ln z^2 M^2}{z^2} + \dots
\er

Where $\pa_{\l} I_{\l \n \; R}$ is the traceless part, that is finite. As we can see, both results agree.

Although in this example we can perform the calculation with both methods with the same effort, with other integrals the situation is different, being necessary to study each expression in order to choose the best one. The explicit evaluation of all the integrals is presented in appendix \ref{ap_integrales}.

\section{Abelian examples}
The two loop renormalization of QED was carried out in \cite{Haagensen:1992vz}, whereas SuperQED was studied in \cite{Song}. We will reobtain the two loop beta function coefficient of those theories using the procedure previously sketched, so that no Ward identity will be used. 

We will make the calculations in both cases in a background field approach  \cite{Abbott:1980hw}, which implies that we will only treat the background field selfenergy (it is the only function needed to be considered when obtaining the beta function \cite{Abbott:1980hw}).

\subsection{QED}
In this calculation, we apply the definitions and conventions of  \cite{Haagensen:1992vz}. The only difference is the background splitting in the gauge field of the form $A_{\m} \rightarrow A_{\m} + B_{\m}$ (similar to the Yang-Mills case \cite{Abbott:1980hw}).
\subsubsection{One loop}

For completeness, we briefly review the CDR renormalization of the photon selfenergy $\Pi_{\m \n}$ \cite{delAguila:1997kw}. Although we need this amplitude to obtain the one loop coefficient of the beta function, it will be of no use when obtaining the two loop one. The bare vacuum polarization is
\br
\Pi_{\m \n}^{(1 \; loop)} &=& - (i e)^2 Tr \left[ \g_{\m} \g^{\l} \pa^x_{\l} \D \g_{\n} \g^{\eps} \pa^y_{\eps} \D \right].
\er

Simplifying the Dirac matrices and applying afterwards CDR we arrive to the renormalized expression
\br
\Pi_{\m \n R}^{(1)} (x) &=& - 4 e^2 \left[ \pa_{\m} \pa_{\n} \D^2 - 2 \D \pa_{\m} \pa_{\n} \D - \frac{1}{2} \d_{\m \n} \Box \D^2 + \d_{\m \n} \D \Box \D \right]_R \nonumber \\
&=& - ( \pa_{\m} \pa_{\n} - \d_{\m \n} \Box ) \left[ - \frac{e^2}{12 \pi^2 ( 4 \pi^2 )} \Box \frac{ \ln x^2 M^2}{x^2} - \frac{e^2}{36 \pi^2} \d (x) \right] \label{QED_1loop}
\er

\subsubsection{Two loop}
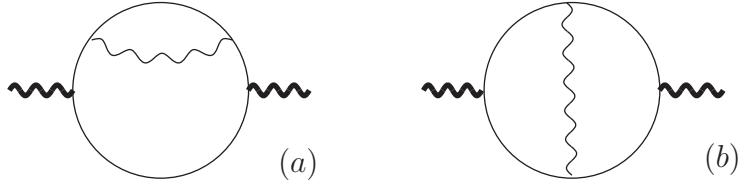
\begin{figure}[t]
\begin{center}
\fcolorbox{white}{white}{
\begin{picture}(293,75) (82,-82)
\SetWidth{0.5}
\SetColor{Black}
\CArc(294,-45)(33.12,119,479)
\SetWidth{2.0}
\Photon(260,-45)(238,-45){2}{3}
\Photon(327,-45)(351,-45){2}{3}
\SetWidth{0.5}
\Photon(292,-12)(294,-77){2}{6}
\CArc(139,-45)(33.12,119,479)
\PhotonArc(139.5,20.64)(53.65,-119.6,-60.4){-2}{4.5}
\SetWidth{2.0}
\Photon(106,-45)(82,-45){2}{3}
\Photon(172,-45)(195,-45){2}{3}
\Text(184,-79)[lb]{\large{\Black{$(a)$}}}
\Text(345,-77)[lb]{\large{\Black{$(b)$}}}
\end{picture}
}
\end{center}
\caption{Two loop QED diagrams}
\label{QED_2loop}
\end{figure}

Proceeding with the two loop case, there are two relevant graphs with external background fields. Diagram (a) has the divergences nested, whereas diagram (b) has overlapping divergences (figure \ref{QED_2loop}). 

\begin{itemize}
\item \bf{Diagram (a)}
\end{itemize}

The expression for this diagram is 
\br
\Pi^{(2 \; a)}_{\m \n} (x) &=& - (ie)^2 \int d^4 u d^4 v \; Tr \left[ \g_{\m} \g^{\l} (- \pa_{\l}^x \D_{xu}) \S^{(1)} (u-v) \g^{\eps} (- \pa_{\eps}^v \D_{vy}) \g_{\n} \g^{\s} ( - \pa_{\s}^y \D_{yx} ) \right] \nonumber \\ 
\er

where $\S^{(1)}$ is the one loop fermion selfenergy. As the term that takes care of the running of the coupling constant in the RG equation is not relevant when obtaining the two loop beta function \cite{Haagensen:1992vz}, we could restrict ourselves to Feynman gauge. In that gauge, the bare fermion selfenergy is
\br
\S^{(1)}(x) &=& - 2 e^2 \g^{\l} \D \pa_{\l} \D (x) 
\er

As we stated in the previous section, CDR imposes a strict order to the operations of index contraction and renormalization: First all the indices should be contracted, and only after that we can renormalize. Inserting the bare fermion selfenergy we are keeping this order here.  

Applying the Clifford algebra of the $\g$-matrices and the integral indentity 
\br
\int d^4 u d^4 v \; \D_{xu} \D_{yv} ( \D_{uv} \pa^u_{\l} \D_{uv} ) = \frac{1}{2} \pa_{\l}^x \int d^4 u d^4 v \; \D_{xu} \D_{yv} \D^2_{uv} 
\er
is straightforward to write $\Pi^{(2 \; a)}_{\m \n}$ in terms of $I^1$ as
\br
\Pi^{(2 \; a)}_{\m \n } (x) &=& 8 e^4 ( \pa_{\m} \D ) ( \pa_{\n} I^1 ) - 4 e^4 \d_{\m \n} ( \pa_{\l} \D ) ( \pa_{\l} I^1 ) 
\er

We only have to replace the different structures made up with $I^1$ by their renormalized values as were listed in section \ref{Nested_div}. The final renormalized expression is then
\br
\Pi^{(2 \; a)}_{\m \n \; R} (x)&=& \frac{e^4}{24(4 \pi^2)^3} \left[ \pa_{\m} \pa_{\n} \frac{- \ln^2 x^2 M^2 - \frac{5}{3} \ln x^2 M^2 }{x^2} + \d_{\m \n} \Box \Box \frac{\ln^2 x^2 M^2 + \frac{8}{3} \ln x^2 M^2}{x^2} \right] + \ldots \nonumber \\
\er

\begin{itemize}
\item \bf{Diagram (b)}
\end{itemize}
This diagram, opposite of the previous one, has overlapping divergences. We will express this in terms of the listed overlapping integrals. The bare expression is
\br
\Pi_{\m \n}^{(2 \; b)} (x-y) &=& - ( i e )^4 \int d^4 u d^4 v Tr \left[ \g_{\m} ( - \g^{\a} \pa^x_{\a} \D_{xu} ) \g^{\rho} ( - \g^{\b} \pa_{\b}^u \D_{uy} ) \g_{\n} ( - \g^{\l} \pa_{\l}^y \D_{yv} ) \g_{\rho} ( - \g^{\s} \pa_{\s} \D_{vx}) \D_{uv} \right] \nonumber \\
\er

Applying algebra of Dirac matrices we can expand this expression, the complete result being listed in appendix \ref{ap_pi_mu_nu}. As can be seen there, the contributions with the delta ($\Box \D = - \d$) can be easily expressed in terms of $I^1$, and its renormalization is as in the previous case. The other contributions can be found in the list of renormalized expressions with overlapping divergences (or can be easily expressed in terms of integrals of the list). So that, we only have to apply directly the results found in section \ref{2loop_CDR}. No Ward identity is needed to fix the correct scale. We found the divergent part of the renormalized expression to be
\br
\Pi^{(2 \; b)}_{\m \n R} (x) &=& - \frac{e^4}{12 (4 \pi^2)^3} \left[ \pa_{\m} \pa_{\n} \Box \frac{ - \ln^2 x^2 M^2 - \frac{14}{3} \ln x^2 M^2}{x^2} + \d_{\m \n} \Box \Box \frac{ \ln^2 x^2 M^2 + \frac{17}{3} \ln x^2 M^2}{x^2} \right] + \ldots \nonumber \\
\er

\begin{itemize}
\item \bf{Final expression}
\end{itemize}

With the two previous results, the total two loop renormalized background selfenergy is
\br
\Pi_{\m \n \; R}^{(2)} (x) &=& 2 \Pi_{\m \n \; R}^{(2 \; a)} + \Pi_{\m \n \; R}^{(2 \; b)} \nonumber \\
&=& \frac{e^4}{4(4 \pi^2)^3} ( \pa_{\m} \pa_{\n} - \d_{\m \n} \Box ) \Box \frac{ \ln x^2 M^2}{x^2} + \ldots
\er

where again $\ldots$ stands for the local terms that we are not taking into account. We can use this result (along with the one loop renormalized value (\ref{QED_1loop})) to obtain straightforwardly as in \cite{Haagensen:1992vz} the two loop expansion of the beta function in terms of $ \a = \frac{e^2}{4 \pi} $ as 
\br
\b ( \a ) = \frac{ 2 \a^2}{3 \pi} + \frac{ \a^3}{2 \pi^2} + {\cal{O}}(\a^{4})
\er

To stress the key points of our calculation, let us compare this procedure with usual differential renormalization \cite{Haagensen:1992vz}. Being $M_{\S}$ and $M_{V}$ the one loop renormalization scales of the fermion selfenergy and the three point vertex $V_{\m}$ respectively, the Ward identity 
\br
\frac{\pa}{\pa z^\m} V_{\m} (x-z,y-z) = - i e \left[ \d^{(4)} (z-x) - \d^{(4)}(z-y) \right] \S (x-y)
\er 

imposes that these scales are related as \cite{Haagensen:1992vz}
\br
\ln \frac{M^2_{\S}}{M^2_{V}} = \frac{1}{2} \label{QED_mass_rel}
\er

When dealing with the two loop contributions, in each case we have to make use of the corresponding one loop scale ($M_{\S}$ or $M_{V}$), being found the final values for $\Pi_{\m \n}^{(2a)}$ and $\Pi_{\m \n}^{(2b)}$ to be
\br
\Pi_{\m \n \; R}^{(2a)} (x)  &=& - \frac{1}{96 \pi^2} \left( \frac{\a}{\pi} \right)^2 \left[ ( \pa_{\m} \pa_{\n} - \d_{\m \n} \Box ) \Box \left( \frac{ \ln^2 x^2 M^2_{\S} + \frac{5}{3} \ln x^2 M^2}{x^2}\right) - \d_{\m \n} \Box \Box \frac{\ln x^2 M^2}{x^2}\right] \nonumber \\
\Pi_{\m \n \; R}^{(2b)} (x) &=& - \frac{1}{48 \pi^2} \left( \frac{\a}{\pi} \right)^2 \left[ - ( \pa_{\m} \pa_{\n} - \d_{\m \n} \Box) \Box \left( \frac{\ln^2 x^2 M^2_V + \frac{17}{3} \ln x^2 M^2}{x^2}\right) + \d_{\m \n} \Box \Box \frac{\ln x^2 M^2}{x^2} \right] \nonumber \\
\er

Then, to obtain the entire two loop vacuum polarization we have to use the mass relation (\ref{QED_mass_rel}) to put one of the scales in terms of the other. As can be seen, all of this is avoided in our procedure. No Ward identity was needed to be imposed, because CDR at one loop has fixed all the ambiguity relevant to logarithms of the scale at the two loop level.

\subsection{SuperQED}

We will now study another example of the use of one loop CDR results in a two loop calculation. In this case, we will discuss the supersymmetric extension of the previous case, SuperQED. This example was already studied in the context of differential renormalization in \cite{Song}, although as in the previous case, Ward identities were imposed to obtain the correct scale relations. In this calculation, we will use the supersymmetric conventions for the $N=1$ superspace of \cite{Gates:1983nr}. With them, the propagator is formed by a product of the delta function of Grassmanian variables and the usual part $P_{ij} = \d (\theta_i - \theta_j) \D (x_{i} - x_{j})$.  The action, expressed in terms of a chiral abelian superfield $W_{\a} = i \bar{D}^2 D_{\a} V$ (where $D_{\a}$ is the supercovariant derivative and $V$ a real superfield ) and chiral matter superfields $\P_{+}$ and $\P_{-}$ is of the form
\br
S &=& \int d^4 x d^2 \th \; W^2 + \int d^4 x d^4 \th \; \bar{\P}_{+} e^{gV} \P_{+} + \int d^4 x d^4 \th \; \bar{\P}_{-} e^{gV} \P_{-}
\er

In this case the background splitting is defined as $ V \rightarrow V + B$ \cite{Song}.

A complete study of perturbative calculations in superspace can be found in \cite{Gates:1983nr}. The basics of the method is to express the diagram in terms of superpropagators and supercovariant derivatives acting on them. Integrating by parts, we have to make the derivatives act on some propagators and afterwards apply the property $\d_{12} \bar{D}^2_1 D^2_1 \d_{12} = \d_{12}$. Here we will write directly the expression found after this step, being the previous standard superspace calculation found in \cite{Song} and \cite{Gates:1983nr} for the SQED case.

\subsubsection{One loop}

For completeness, as in the QED case, we consider the one loop renormalization, although we will not  use it in the two loop calculation. The bare amplitude is
\br
\G^{(1 loop)} &=& \frac{g^2}{4} \int d^4 x d^4 y d^4 \th B (x,\th) \left( D^{\a} \bar{D}^2 D_{\a} B (y, \th) \right) \D^2_{xy} + \nonumber \\
&+& \frac{g^2}{2} \int d^4 x d^4 y d^4 \th B (x, \th) B ( y, \th) \D_{xy} \Box \D_{xy} 
\er

and renormalizing according to CDR rules we found 
\br
\G^{(1 \; loop)}_R  &=& - \frac{g^2}{16 (4 \pi^2)^2} \int d^4 x d^4 y d^4 \th \; B (x, \th) \left( D^{\a} \bar{D}^2 D_{\a} B (y, \th) \right) \Box \frac{ \ln (x-y)^2 M^2 }{(x-y)^2} \nonumber \\
\er

\subsubsection{Two loops}

\begin{figure}[t]
\begin{center}
\fcolorbox{white}{white}{
\begin{picture}(286,200) (107,-81)
\SetWidth{0.5}
\SetColor{Black}
\Photon(163,-8)(166,-73){2}{6}
\CArc(166,-40)(33.12,119,479)
\SetWidth{2.0}
\Photon(132,-40)(110,-40){2}{3}
\Photon(199,-40)(223,-40){2}{3}
\SetWidth{0.5}
\CArc(164,80)(33.12,119,479)
\PhotonArc(164.5,145.5)(53.52,-119.68,-60.32){-2}{4.5}
\SetWidth{2.0}
\Photon(131,80)(107,80){2}{3}
\Photon(197,80)(217,80){2}{3}
\SetWidth{0.5}
\PhotonArc(335.33,108.25)(28.73,170.48,293.97){-2}{4.5}
\CArc(315,81)(33.12,119,479)
\SetWidth{2.0}
\Photon(282,81)(258,81){2}{3}
\Photon(348,81)(371,81){2}{3}
\SetWidth{0.5}
\CArc(316,-39)(33.12,119,479)
\SetWidth{2.0}
\Photon(284,-39)(262,-39){2}{3}
\Photon(349,-39)(374,-39){2}{3}
\SetWidth{0.5}
\Photon(284,-39)(349,-39){2}{6}
\Text(363,-80)[lb]{\large{\Black{$(d)$}}}
\Text(359,39)[lb]{\large{\Black{$(b)$}}}
\Text(208,38)[lb]{\large{\Black{$(a)$}}}
\Text(211,-82)[lb]{\large{\Black{$(c)$}}}
\end{picture}
}
\end{center}
\caption{Two loop SQED diagrams}
\label{SQED_2loop}
\end{figure}
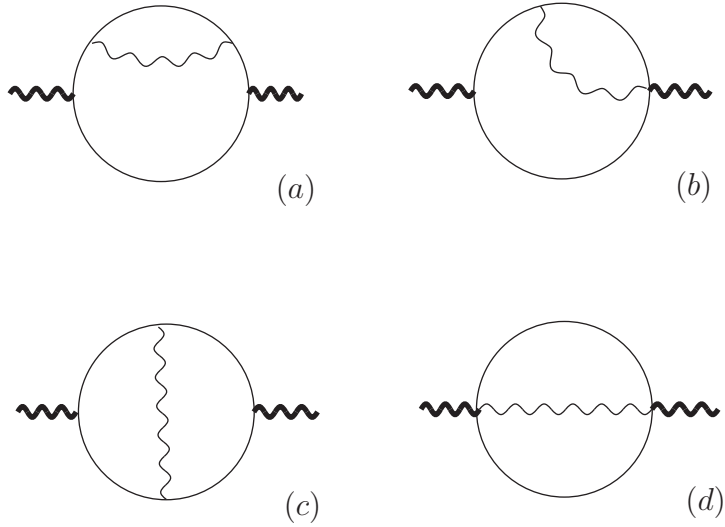


We first list the bare expressions of diagrams presented in figure \ref{SQED_2loop}. We will write these expressions in terms of $I^1$ and integral (\ref{int9}). Then, we renormalize them according to the procedure defined in section \ref{2loop_CDR}.

\begin{itemize}
\item \bf{Diagram (a)}
\end{itemize}
\br
\G^{(2 \; loop)}_a &=& \frac{g^4}{2} \int d^4 x d^4 y d^4 \th \; B (x, \th) B(y, \th) \left[ \Box (\D I^1) - 2 \D^3 - \pa^{\a \dot{\a}} ( \D \pa_{\a \dot{\a}} I^1 ) \right] + \nonumber \\
&+& \frac{g^4}{2} \int d^4 x d^4 y d^4 \th \; B(x, \th) \left( D^{\a} \bar{D}^2 D_{\a} B (y, \th) \right) \D I^1
\er

\begin{itemize}
\item \bf{Diagram (b)}
\end{itemize}
\br
\G^{(2 \; loop)}_b &=& g^4 \int d^4 x d^4 y d^4 \th B (x, \th) B (y, \th) \left[ - \Box ( \D I^1 ) + 2 \D^3 + \pa^{\a \dot{\a}} ( \D \pa_{\a \dot{\a}} I^1 ) \right] - \nonumber \\
&-& g^4 \int d^4 x d^4 y d^4 \th \; B (x,\th) \left( D^{\a} \bar{D}^2 D_{\a} B (y, \th) \right) \D I^1
\er

\begin{itemize}
\item \bf{Diagram (c)}
\end{itemize}
\br
\G^{(2 \; loop)}_c &=& \frac{g^4}{2} \int d^4 x d^4 y d^4 \th B ( x, \th) \left( D^{\a} \bar{D}^2 D_{\a} B(y, \th) \right) \left[ \D I^1 \right] + \nonumber \\
&+& \frac{g^4}{2} \int d^4 x d^4 y d^4 \th \; B (x, \th) B(y, \th) \left[ \Box ( \D I^1 ) - \D^3 - \pa^{\a \dot{\a}} ( \D \pa_{\a \dot{\a}} I^1) \right] + \nonumber \\
&+& \frac{g^4}{2} \int d^4 x d^4 y d^4 \th \; B(x, \th) \left( D^{\b} \bar{D}^2 D^{\a} B(y, \th) \right) C^{\dot{\b} \dot{\a}} \; H[ \pa_{\b \dot{\b}},1 \; ; \; 1, \pa_{\a \dot{\a}}] \nonumber \\
\label{SQED_diag_e}
\er

Where $C^{\dot{\a} \dot{\b}}$ is the raising and lowering matrix of the SU(2) supersymmetric indices.

\begin{itemize}
\item \bf{Diagram (d)}
\end{itemize}
\br
\G^{(2 \; loop)}_d &=& - \frac{g^4}{2} \int d^4 x d^4 y d^4 \th \; B(x, \th) B(y, \th) \D^3 
\er

\begin{itemize}
\item \bf{Final renormalized expression}
\end{itemize}

We can now substitute the previous expressions by its renormalized values and add them in the final result. It can be easily shown that only the the last part of (\ref{SQED_diag_e}) survives, and the rest cancels exactly. But we are allowed to do this only because we are using CDR at one loop. In \cite{Song} this simplified calculation was forbidden. One has to renormalize in a separate form each diagram; obtain the renormalized results with different scales and relate them via Ward identities. Only at this point we can add up the expressions. With our method, we always renormalize the same structures in the same way and with the same scale, so that we can add up the expressions even before performing the renormalization. 

The final result is then (multiplying by 2 because we have performed the calculations with only one of the chiral matter fields $\P_{+}$, $\P_{-}$).
\br   
\G^{(2 \; loop)}_R &=&  2 \left ( \G^{(2)}_{a \; R} + \G^{(2)}_{b \; R} + \G^{(3)}_{c \; R} + \G^{(4)}_{d \; R} \right) \nonumber \\
&=& g^4 \int d^4 x d^4 y d^4 \th \; B(x, \th) \left[ D^{\b} \bar{D}^2 D^{\a} B (y, \th)\right] C^{\dot{\b} \dot{\a}}  \; H^R[ \pa_{\b \dot{\b}},1 \; ; \; 1, \pa_{\a \dot{\a}}] \nonumber \\
&=& - \frac{g^4}{16 (4 \pi^2)^3} \int d^4 x d^4 y d^4 \th B (x, \th) \left( D^{\a} \bar{D}^2 D_{\a} B (y, \th) \right) \Box \frac{ \ln (x-y)^2 M^2}{(x-y)^2} + \ldots \nonumber \\
\er

Where we have used the result (\ref{int9}) of the list of integrals with overlapping divergences. Taking into account that with our conventions (those of reference \cite{Gates:1983nr})  $g = \sqrt{2} g_{SQED}$, with $g_{SQED}$ the usual coupling constant of SuperQED, we can easily evaluate the two loop expansion of the beta function as in \cite{Song} to be
\br
\b( g_{SQED} ) &=& \frac{1}{8 \pi^2} g^3_{SQED} + \frac{1}{32 \pi^4} g^5_{SQED} + {\cal{O}}(g^7_{SQED})
\er 

Which agrees with previous results found in the literature \cite{Shifman:1985fi,Novikov:1985rd}.

\section{Non abelian example}

One of the relevant calculations that was not yet obtained with differential renormalization is the two loop renormalization of Yang-Mills theory. Here we will detail how we can obtain the two loop coefficient of the beta function of this theory with little effort, making use of the procedure defined in section \ref{2loop_CDR}.

We will follow \cite{Abbott:1980hw} where a two loop calculation of the beta function of Yang-Mills theory in the background field approach with dimensional renormalization was performed. The Yang-Mills lagrangian, written in terms of gauge covariant derivatives ($D_{\m}^{ac} = \pa_{\m} \d^{ac} + g f^{abc} B_{\m}^b $ and ${\bf{D}}_{\m}^{ac} = \pa_{\m} \d^{ac} + g f^{abc} ( B_{\m}^a + A_{\m}^a)$) and ghost fields $\eta^a$ is
\br
\cal{L} &=& \frac{1}{4} F^a_{\m \n} F^a_{\m \n} + \frac{1}{2 \a} ( D_{\m} A_{\m} )^a ( D_{\n} A_{\n} )^a + ( D_{\m} \bar{\eta})^a ( \bf{D}_{\m} \eta )^a \;.
\er

We have to point out that the gauge fixing parameter $\a$ will be redefined in our calculation as $\frac{1}{\a} = 1 + \xi$, so that Feynman gauge ($\a = 1$) will correspond to $\xi = 0$.

\subsection{One loop renormalization}

We now briefly review the results found in \cite{Perez-Victoria:1998fj} and in \cite{Freedman:1991tk} for the one loop differential renormalization of the quantum and background fields. Although the background selfenergy is all that we need when obtaining the one loop beta function, we also have to consider the renormalization of the quantum selfenergy as this will be a one loop insertion in a two loop diagram (diagram (b)). Also, it will be needed in order to obtain the first coefficient of the expansion of the function that takes into account the running of the gauge parameter in the renormalization group equation. We will detail this later in section \ref{sec_RG}.

\subsubsection{Correction to the $B_{\m}^a$ propagator  \\ (Feynman gauge $\xi=0$)}

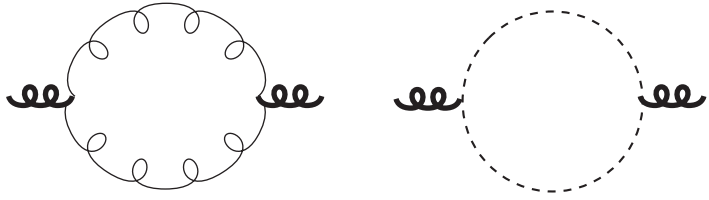
\begin{figure}
\begin{center}
\fcolorbox{white}{white}{
\begin{picture}(262,88) (5,-1)
\SetWidth{0.8}
\SetColor{Black}
\DashCArc(210,40)(33.94,135,495){3}
\SetWidth{2.0}
\Gluon(29,42)(5,42){3.5}{2.57}
\Gluon(175,41)(151,41){3.5}{2.57}
\Gluon(267,42)(243,42){3.5}{2.57}
\Gluon(123,42)(99,42){3.5}{2.57}
\SetWidth{0.5}
\GlueArc(64.5,37.2)(34.83,7.92,172.08){5}{4.29}
\GlueArc(64,47.4)(35.41,-171.23,-8.77){5}{4.29}
\end{picture}
}
\end{center}
\caption{One loop YM diagrams}
\label{YM_1loop}
\end{figure}

This is the sum of two different diagrams: one with a loop of quantum gauge fields, and another with ghost fields (see figure \ref{YM_1loop}).

The procedure is as defined in the CDR section. First we write the expressions in terms of the basic functions, and after that we replace them by their renormalized values.

\begin{itemize}
\item {\bf Gauge loop}
\end{itemize}

Here and in the rest of the following diagrams of Yang-Mills theory, $D_{\m}^{x,y}$ denotes a space-time derivative acting on one external field. Applying the Leibniz rule, this becomes a minus derivative acting on the propagators.
\br
<B_{\m}^a (x) B_{\n}^b (y) > &=& \frac{ g^2 f^{acd} f^{bdc}}{2} \D_{xy} \left[ - 2 \d_{\m \s} D^x_{\rho} + \d_{\rho \s} (\stackrel{\leftarrow}{\pa_{\m}^{x}} - \pa_{\m}^x) + 2 \d_{\m \rho} D_{\s}^x \right] \times \nonumber \\
&\times& \left[ - 2 \d_{\n \rho} D_{\s}^y + \d_{\rho \s} (\pa_{\n}^y - \stackrel{\leftarrow}{\pa_{\n}^y}) + 2 \d_{\n \s} D^y_{\rho} \right] \D_{xy} \nonumber \\&=& \frac{g^2 C_A \d^{ab}}{2} \left[ 8 \pa_{\m} \pa_{\n} \D^2 - 8 \d_{\m \n} \Box \D^2 + 8 \pa_{\m}( \D \pa_{\n} \D ) - 16 \D \pa_{\m} \pa_{\n} \D \right] \nonumber \\
\er

\begin{itemize}
\item {\bf Ghost loop}
\end{itemize}
\br
<B_{\m}^a (x) B_{\n}^b (y) > &=& - g^2 f^{abc} f^{bcd} \D_{xy} ( \stackrel{\leftarrow}{\pa_{\m}^{x}} - \pa_{\m}^x) (\pa_{\n}^y - \stackrel{\leftarrow}{\pa_{\n}^y}) \D_{xy} \nonumber \\
&=& - g^2 C_A \d^{ab} \left[ 2 \pa_{\m} ( \D \pa_{\n} \D ) - 4 \D \pa_{\m} \pa_{\n} \D \right] \nonumber \\
\er

\begin{itemize}
\item {\bf Total contribution}
\end{itemize}
The total non-renormalized contribution is
\br
<B_{\m}^a(x) B_{\n}^b(y)> &=& g^2 C_A \d^{ab} \left[ 4 \pa_{\m} \pa_{\n} \D^2 - 4 \d_{\m \n} \Box \D^2 + 2 \pa_{\m} ( \D \pa_{\n} \D ) - 4 \D \pa_{\m} \pa_{\n} \D \right] 
\er

Replacing the values prescribed by CDR, the renormalized one loop contribution to the $B_{\m}^a$ propagator is obtained as
\br
<B_{\m}^a(x) B_{\n}^b(0)>_R &=&  g^2 C_A \d^{ab} (\pa_{\m} \pa_{\n} - \d_{\m \n} \Box) \left[ \frac{11}{3} \D^2_R - \frac{1}{72 \pi^2} \d (x) \right] \nonumber \\
&=& g^2 C_A \d^{ab} (\pa_{\m} \pa_{\n} - \d_{\m \n} \Box) \left[ - \frac{11}{48 \pi^2 (4 \pi^2)} \Box \frac{\ln x^2 M^2}{x^2} - \frac{1}{72 \pi^2} \d (x) \right]  \nonumber \\ \label{1_loop}
\er

The result found here is transverse, fulfilling the corresponding Ward identity as was guaranteed by the use of CDR.

\subsubsection{Correction to the $A_{\m}^a$ propagator  \\ (Feynman gauge $\xi=0$)}

As in the previous case, we have contributions with gauge and ghost loops. Proceeding in the same way as in the background selfenergy, we first obtain the full expanded bare expressions and then renormalize them according to CDR rules.

\begin{itemize}
\item {\bf Gauge loop}
\end{itemize}
\br
<A_{\m}^a (x) A_{\n}^b (y) > &=& \frac{g^2 f^{acd} f^{bdc}}{2} \D_{xy} \left[ \d_{\m \rho} ( D_{\s}^x - \stackrel{\leftarrow}{\pa_{\s}^{ x}}) + \d_{\s \m} ( \pa_{\rho}^x - D_{\rho}^x ) + \d_{\rho \s} ( \stackrel{\leftarrow}{\pa_{\m}^{x}} -\pa_{\m}^x ) \right] \times \nonumber \\
& & \times \left[ \d_{\n \s} (D_{\rho}^y - \pa_{\rho}^y) + \d_{\rho \n} ( \stackrel{\leftarrow}{\pa_{\s}^{y}} - D_{\s}^y) + \d_{\rho \s} ( \pa_{\n}^y - \stackrel{\leftarrow}{\pa_{\n}^{y}} ) \right] \D_{xy} \nonumber \\
&=& \frac{g^2 C_A \d^{ab}}{2}  \left[ 2 ( \pa_{\m} \pa_{\n} - \d_{\m \n} \Box ) \D^2 + 10 \pa_{\m} ( \D \pa_{\n} \D ) - 10 \D \pa_{\m} \pa_{\n} \D -  \right. \nonumber \\
&-& \left. 4 \d_{\m \n} \pa^{\l} ( \D \pa_{\l} \D ) - 2 \d_{\m \n} \D ( \Box \D ) \right] 
\er

\begin{itemize}
\item {\bf Ghost loop}
\end{itemize}
\br
<A_{\m}^a (x) A_{\n}^b (y) > &=& - g^2 f^{adc} f^{bcd} \D_{xy} \stackrel{\leftarrow}{\pa_{\m}^{ x}} \pa_{\n}^y \D_{xy} \nonumber \\
&=& - g^2 C_A \d^{ab} \left[ \pa_{\m}( \D \pa_{\n} \D) - \D \pa_{\m} \pa_{\n} \D \right] 
\er

\begin{itemize}
\item {\bf Total contribution}
\end{itemize}

Adding the two previous results we found the non-renormalized contribution to be\br
<A_{\m}^a (x) A_{\n}^b (y) > &=& g^2 C_A \d^{ab} \left[ \pa_{\m} \pa_{\n} \D^2 - \d_{\m \n} \Box \D^2 + 4 \pa_{\m} ( \D \pa_{\n} \D ) - 2 \d_{\m \n} \pa^{\l} ( \D \pa_{\l} \D ) - \right. \nonumber \\
&-& \left. 4 \D \pa_{\m} \pa_{\n} \D - \d_{\m \n} \D ( \Box \D ) \right] 
\er

and with CDR identities it is straightforward to obtain the renormalized contribution as
\br
<A_{\m}^a (x) A_{\n}^b (0) >_R &=& g^2 C_A \d^{ab} \left[ \frac{5}{3}(  \pa_{\m} \pa_{\n} - \d_{\m \n } \Box) \D^2_R - \frac{1}{72 \pi^2} (\pa_{\m} \pa_{\n} - \d_{\m \n} \Box) \d (x) \right] \nonumber \\
&=& - \frac{g^2 C_A \d^{ab}}{144 \pi^2} (\pa_{\m} \pa_{\n} - \d_{\m \n} \Box) \left[ \frac{15}{4 \pi^2} \Box \frac{ \ln x^2 M^2}{x^2} + 2 \d (x) \right] \label{1loop_quantum}
\er

\subsubsection{Effective action in a generic gauge}

All of this calculation is performed in Feynman gauge, so that we have to take care of the running of the gauge parameter. In order to do that, we consider a functional approach. 

The idea is to expand the effective action at one loop at second order in the background fields and retain only the linear dependence in the gauge parameter, due to the fact that in the renormalization group equation we will derive wrt. this parameter and after that impose Feynman gauge ($\xi$=0). 

To obtain the exact one loop effective action it is well known that we have to consider only the part of the lagrangian quadratic in the $A_{\m}^a$ fields. This part is
\br
{\cal{L}}^{(2)}_{gauge} &=& g f^{abc} B_{\m \n}^a A_{\m}^b A_{\n}^c + \frac{1}{2} (D_{\m} A_{\n})^a (D_{\m} A_{\n})^a + \frac{\xi}{2} (D_{\m} A_{\m})^a (D_{\n} A_{\n})^a \nonumber \\
&=& - \frac{1}{2} A_{\m}^a \left[ \d_{\m \n} \Box^{ab} - 2 g f^{cab} B_{\m \n}^c + \xi (D_{\m} D_{\n} )^{ab} \right] A_{\n}^b 
\er

where ${\Box}^{ab} = ( D^{\m} D_{\m})^{ab} $. 

Then, the generating functional for connected Green functions can be put as
\br
W &=& - \frac{1}{2} tr \ln \left[ \d_{\m \n} \Box^{ab} - 2 g f^{cab} B_{\m \n}^c + \xi (D_{\m} D_{\n})^{ab} \right] 
\er

At first order in $\xi$ and second order in the $B$ fields, this can be expressed as
\br
W &=&  \xi C_A g^2 \d^{cg} tr \left[ \frac{1}{2} \Box^{-1} B_{\m \n}^c \Box^{-1} B_{\m \n}^g + 2 (\Box^{-1}) B_{\m \n}^c (\Box^{-1}) B_{\n \l}^g (\Box^{-1}) \pa_{\l} \pa_{\m} \right] \nonumber \\
&=& \xi C_A g^2 tr \left[ \frac{1}{2} \D B_{\m \n}^a \D B_{\m \n}^a - 2 \D B_{\m \n}^a \D B_{\n \l}^a \D \pa_{\l} \pa_{\m} \right] \label{effective_1loop}
\er

where as usual $\Box = \pa^{\m} \pa_{\m}$.

We can write the renormalized expression of the first term of (\ref{effective_1loop}) as
\br
(A) &=& \frac{1}{2} \int d^4 x d^4 y \; B_{\m \n}^a (x) B_{\m \n}^a (y)  \D^2 |_R \nonumber \\
\er

Whereas the second one is of the following form
\br
(B) &=& - 2 \int d^4 x d^4 y d^4 u \; ( \pa_{\l}^u \pa_{\m}^u \D_{ux}) B_{\m \n}^a (x) B_{\n \l}^a (y) \D_{xy} \D_{yu} \nonumber \\
&=& - 2 \int d^4 x d^4 y \; B_{\m \n}^a (x) B_{\n \l}^a(y) \D_{xy} \pa_{\l}^x \pa_{\m}^x \int d^4 u \; \D_{xu} \D_{yu}
\er

In order to evaluate this expression we must apply CDR in momentum space, arriving to
\br
\int d^4 u \; \D_{xu} \D_{uy} &=& - \frac{1}{4(4 \pi^2)} \ln (x-y)^2 m^2 \nonumber \\
&\equiv& - \bar{\D} (x-y)
\er

\br
(B) &=& 2 \int d^4 x d^4 y \; B_{\m \n}^a (x) B_{\n \l}^a (y) \left( \D_{xy} \pa_{\l} \pa_{\m} \bar{\D} \right)
\er

Remembering the CDR renormalization of $\D \pa_{\l} \pa_{\m} \bar{\D}$ listed in (\ref{CDR_gen_gauge}) we can obtain the final renormalized expression for $(B)$ as
\br
(B) &=& - \frac{1}{2} \int d^4 x d^4 y \; B_{\m \n}^a (x) B_{\m \n}^a (y) \D^2 |_R - \frac{1}{16 \pi^2} \int d^4 x d^4 y B_{\m \n}^a (x) B_{\n \l}^a (y) \pa_{\m} \pa_{\l} \D  \nonumber \\
\er

Adding up the two results
\br
(A)+(B) &=& - \frac{\xi C_A g^2}{4(4 \pi^2)} \int d^4 x d^4 y \; B_{\m \n}^a B_{\n \l}^a \pa_{\m} \pa_{\l} \D
\er

This can be written at explicit second order in $B$ fields as
\br
\int B_{\m \n}^a (x) B_{\n \l}^a \pa_{\m} \pa_{\l} \D &=&  \int (\pa_{\m} B_{\n}^a - \pa_{\n} B_{\m}^a )(\pa_{\n} B_{\l}^a - \pa_{\l} B_{\n}^a ) \pa_{\m} \pa_{\l} \D +{\cal{O}}(B^3) \nonumber \\
&=& - \int B_{\m}^a (x) B_{\n}^a (y) ( \pa_{\m} \pa_{\n} - \d_{\m \n} \Box) \Box \D + {\cal{O}}(B^3) \nonumber \\
\er

Finally, remembering $\G = -W + \int J \phi$ we can obtain the linear dependence in the gauge parameter $\xi$ of the contribution at one loop in a generic gauge to the background two point function as
\br
\G_\xi = - \frac{\xi C_A g^2}{4(4 \pi^2)} \int d^4 x d^4 y \; B_{\m}^a (x) B_{\n}^a (y) (\pa_{\m} \pa_{\n} - \d_{\m \n} \Box) ( \Box \D ) \label{gauge_fix_ren}
\er

\subsection{Two loop diagrams}

\begin{figure}[t]
\begin{center}
\fcolorbox{white}{white}{
\begin{picture}(341,364) (0,0)
\SetWidth{0.5}
\SetColor{Black}
\GlueArc(98.96,199.1)(30.12,-174.09,-4){4}{5.14}
\SetWidth{2.0}
\Gluon(67,196)(42,196){3.5}{2.57}
\SetWidth{0.5}
\Gluon(69,115)(133,115){4.5}{5.24}
\SetWidth{2.0}
\Gluon(157,114)(133,115){3.5}{2.57}
\SetWidth{0.5}
\GlueArc(101,119.29)(32.29,-172.37,-7.63){4}{4.29}
\GlueArc(261.46,196.85)(33.48,119.44,178.03){4}{2.57}
\GlueArc(249.27,193.78)(45.93,5.27,44.54){4}{2.57}
\GlueArc(102.03,200.03)(28.26,-8.21,58.02){4}{2.57}
\GlueArc(112.73,191.78)(45.93,135.46,174.73){4}{2.57}
\SetWidth{0.8}
\DashCArc(98,293)(31.91,148,508){3}
\DashCArc(262,227)(18.38,135,495){3}
\SetWidth{2.0}
\Gluon(326,115)(302,115){3.5}{2.57}
\Gluon(319,198)(295,198){3.5}{2.57}
\Text(227,6)[lb]{\large{\Black{$(f)$}}}
\SetWidth{0.8}
\DashCArc(182,38)(35.36,135,495){3}
\SetWidth{2.0}
\Gluon(147,36)(123,37){3.5}{2.57}
\Gluon(240,38)(217,38){3.5}{2.4}
\SetWidth{0.5}
\GlueArc(137.26,80.27)(45.32,-77.59,-9.23){4}{4.29}
\SetWidth{2.0}
\Gluon(69,115)(42,115){3.5}{2.57}
\Gluon(226,115)(202,115){3.5}{2.57}
\Gluon(159,196)(130,196){3.5}{2.57}
\SetWidth{0.5}
\GlueArc(101,110.73)(32.28,7.6,172.4){4}{5.14}
\GlueArc(98.5,218.79)(19.22,15.74,164.26){4}{4.29}
\GlueArc(98.5,229.21)(19.22,-164.26,-15.74){4}{4.29}
\GlueArc(245,115.26)(19,-0.8,180.8){4}{4.29}
\GlueArc(245,116.03)(19.03,-176.9,-3.1){4}{4.29}
\GlueArc(283,112.88)(19.12,6.36,173.64){4}{4.29}
\GlueArc(283,115.33)(19,-178.99,-1.01){4}{4.29}
\GlueArc(261.5,206.46)(34.55,-165.82,-14.18){4}{5.14}
\SetWidth{2.0}
\Gluon(228,198)(203,198){3.5}{2.74}
\Gluon(66,292)(42,293){3.5}{2.57}
\SetWidth{0.8}
\DashLine(229,292)(291,293){3}
\SetWidth{2.0}
\Gluon(317,293)(291,293){3.5}{2.57}
\SetWidth{0.5}
\GlueArc(260,299.64)(31.7,-167.91,-12.09){4.5}{5.14}
\SetWidth{0.8}
\DashCArc(260,294.95)(31.06,-3.61,183.61){3}
\SetWidth{2.0}
\Gluon(229,293)(206,293){3.5}{2.11}
\Text(300,265)[lb]{\large{\Black{$(c)$}}}
\Gluon(154,293)(130,293){3.5}{2.57}
\SetWidth{0.5}
\GlueArc(98,329.88)(27.95,-145.39,-34.61){4}{4.29}
\Text(134,261)[lb]{\large{\Black{$(a)$}}}
\Text(311,86)[lb]{\large{\Black{$(e)$}}}
\Text(146,84)[lb]{\large{\Black{$(d)$}}}
\Text(181,195)[lb]{\large{\Black{$+$}}}
\Text(182,168)[lb]{\large{\Black{$(b)$}}}
\end{picture}
}
\caption{Two loop YM diagrams a-f}
\label{2loop_1}
\end{center}
\end{figure}
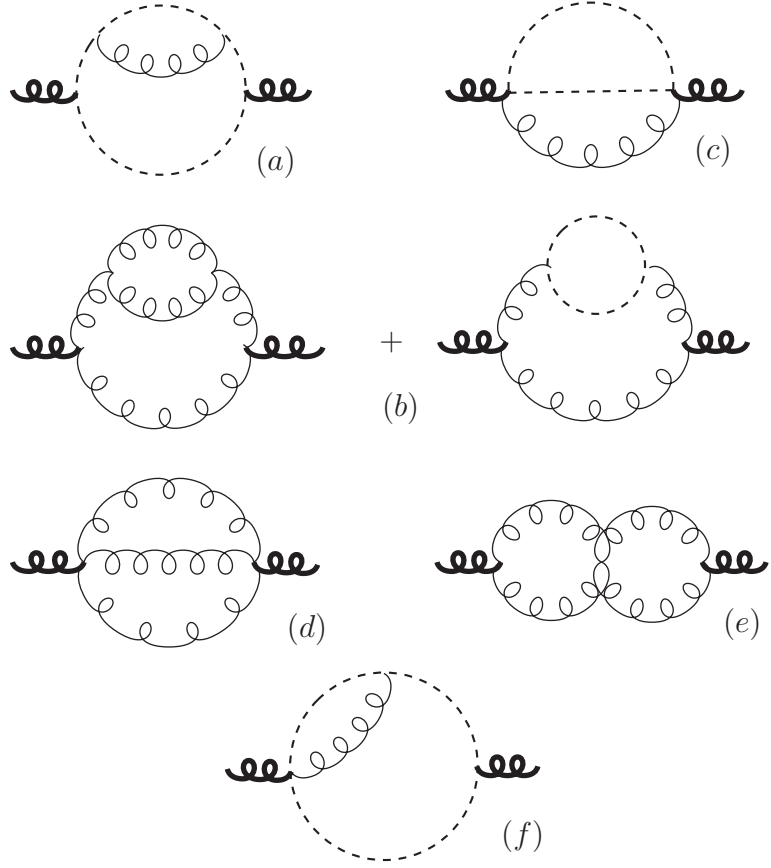

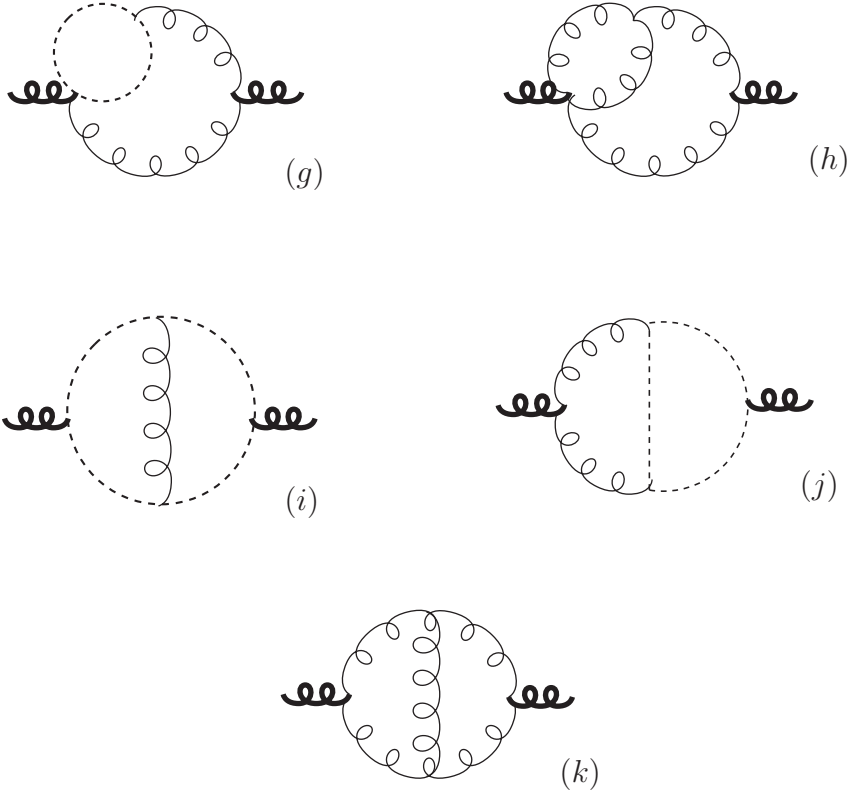
\begin{figure}[t]
\begin{center}
\fcolorbox{white}{white}{
\begin{picture}(334,305) (33,-29)
\SetWidth{0.5}
\SetColor{Black}
\GlueArc(192.97,7.97)(30.04,1.96,176.13){4}{5.14}
\GlueArc(193,12.09)(30.16,-174.11,-5.89){4}{5.14}
\Gluon(192,41)(192,-21){5}{4.29}
\SetWidth{2.0}
\Gluon(163,10)(138,10){3.5}{2.74}
\Gluon(247,9)(223,9){3.5}{2.57}
\Text(243,-26)[lb]{\large{\Black{$(k)$}}}
\SetWidth{0.8}
\DashCArc(92,117)(35.36,135,495){2.6}
\SetWidth{2.0}
\Gluon(57,114)(33,114){3.5}{2.57}
\Gluon(150,114)(126,114){3.5}{2.57}
\SetWidth{0.5}
\Gluon(90,152)(91,81){5}{4.29}
\Text(139,77)[lb]{\large{\Black{$(i)$}}}
\SetWidth{0.6}
\DashCArc(281.18,118.5)(31.92,-99.33,99.33){2.2}
\SetWidth{2.0}
\Gluon(337,121)(313,121){3.5}{2.57}
\Gluon(244,119)(219,119){3.5}{2.74}
\SetWidth{0.5}
\GlueArc(270.29,121.25)(26.38,77.49,184.88){4}{3.43}
\GlueArc(268.52,113.59)(24.13,167.05,290.58){4}{3.43}
\SetWidth{0.6}
\DashLine(276,147)(276,87){2.2}
\Text(334,83)[lb]{\large{\Black{$(j)$}}}
\SetWidth{2.0}
\Gluon(145,237)(119,237){3.5}{2.9}
\Gluon(60,237)(35,237){3.5}{2.74}
\SetWidth{0.5}
\GlueArc(89.46,238.57)(29.58,-175.01,-3.05){4}{5.14}
\GlueArc(90.57,236.9)(28.43,0.21,107.56){4}{3.43}
\SetWidth{0.8}
\DashCArc(70,252)(18.38,135,495){2.2}
\Text(139,201)[lb]{\large{\Black{$(g)$}}}
\SetWidth{2.0}
\Gluon(246,237)(222,237){3.5}{2.57}
\SetWidth{0.5}
\GlueArc(260.42,247.8)(18.82,59.41,215){4}{3.43}
\GlueArc(254.43,253.34)(18.87,-119.99,34.39){4}{3.43}
\GlueArc(277.46,238.57)(29.58,-175.01,-3.05){4}{5.14}
\GlueArc(278.57,236.9)(28.43,0.21,107.56){4}{3.43}
\SetWidth{2.0}
\Gluon(331,237)(307,237){3.5}{2.57}
\Text(337,206)[lb]{\large{\Black{$(h)$}}}
\end{picture}
}
\end{center}
\caption{Two loop YM diagrams g-k}
\label{2loop_2}
\end{figure}

Now we follow with the two loop contribution to the background field selfenergy. Diagrams are those of figures \ref{2loop_1} and \ref{2loop_2} ((a) to (k)). First as an example, we will renormalize diagrams (a) and (k), as these two diagrams are an example of the two different types of divergences that we can find: Diagram (a) has nested divergences, whereas diagram (k) has overlapping divergences. The explicit evaluation of the rest of the diagrams is presented in appendix \ref{ap_two_loop}. 

After discussing in detail those two diagrams, we will list the renormalized expressions of all of the two loop contributions that add up into the two loop renormalized background selfenergy.

\subsubsection{Diagram (a)}

This diagram has the following form
\br
< B_{\m}^a (x) B_{\n}^b (y) >_a &=& - 2 g^4 f^{aec} f^{bcd} f^{gdf} f^{gfe} \int  d^4 u d^4 v \; \D_{xy} ( \stackrel{\leftarrow}{\pa_{\m}^x} - \pa_{\m}^x) ( \pa_{\n}^y - \stackrel{\leftarrow}{\pa_{\n}^y}) \D_{yv} \times \nonumber \\
& & \times ( \pa_{\l}^v \D_{uv}) \D_{uv} ( \pa_{\l}^u \D_{xu}) \nonumber \\ 
\er

We can rearrange this expression in terms of the integral $I^1$ previously defined as 
\br
< B_{\m}^a (x) B_{\n}^b (y) >_a &=& - g^4 C_A^2 \d^{ab} \left[ 4 \pa_{\n} ( \D \pa_{\m} I^1 ) - \pa_{\m} \pa_{\n} ( \D I^1 ) - 4 \D \pa_{\m} \pa_{\n} I^1 \right]
\er

Now, in order to renormalize, we have to substitute these expressions with their renormalized values (obtained in section \ref{Nested_div}), arriving to
\br
<B_{\m}^a(x) B_{\n}^b (0) >_{a\; R} &=& \frac{g^4 C_A^2 \d^{ab}}{32(4 \pi^2)^3} \left[ \pa_{\m} \pa_{\n} \Box \frac{ - \frac{1}{3} \ln^2 x^2 M^2 - \frac{8}{9} \ln x^2 M^2}{x^2} \right. \nonumber \\
& & \left. + \d_{\m \n} \Box \Box \frac{\frac{1}{3}  \ln^2 x^2 M^2 + \frac{11}{9} \ln x^2 M^2}{x^2} \right] + \ldots
\er

\subsubsection{Diagram (k)}

In order to obtain all the contributions that form this diagram, the Mathematica package 'FeynCalc' was used, so that all the index contractions were performed by the computer, being its output the final relevant expressions that need to be renormalized. The contributions shown here are those that have a divergent part, omitting those terms that are finite.
\br
< B_{\m}^a (x) B_{\n}^b (y) >_k = \frac{1}{4} C_A^2 \d^{ab} &\left[\right.& + 16 \d_{\m \n} \pa_{\l}^x \pa_{\s}^x H[1,\pa_{\l} \pa_{\s} \; ; \; 1,1] - 20 \pa_{\n}^x \pa_{\l}^x H[1, \pa_{\m} \pa_{\l} \; ; \; 1,1] - \nonumber \\
& & - 124 \pa_{\n}^x \pa_{\l}^x H [ 1, \pa_{\l} \; ; \; 1, \pa_{\m}] + 72 \pa_{\l}^x H[1, \pa_{\m} \pa_{\l} \; ; \; 1, \pa_{\n}] + \nonumber \\
& & + 56 \d_{\m \n} \pa_{\l}^x \pa_{\s}^x H[1, \pa_{\l} \; ; \; 1, \pa_{\s}] - 72 \pa_{\n}^x H[ 1, \pa_{\m} \Box \; ; \; 1,1] + \nonumber \\
& & + 20 \pa_{\m}^x \pa_{\n}^x H [ 1 , \Box \; ; \; 1, 1] - 144 H[ 1, \pa_{\m} \Box \; ; \; 1, \pa_{\n} ] + \nonumber \\
& & + 72 \pa_{\m}^x H[1, \Box \; ; \; 1, \pa_{\n}] - 72 \pa_{\n}^x H[1, \pa_{\m} \pa_{\l} \; ; \; 1, \pa_{\l}] + \nonumber \\
& & + 34 \pa_{\m}^x \pa_{\n}^x H[ 1, \pa_{\l} \; ; \; 1, \pa_{\l} ] - 72 H[ 1, \pa_{\m} \pa_{\l} \; ; \; 1, \pa_{\n} \pa_{\l} ] + \nonumber \\
& & + 40 \Box H [ 1, \pa_{\m} \pa_{\n} \; ; \; 1, 1] + 32 \Box H [ 1, \pa_{\m} \; ; \; 1, \pa_{\n}] + \nonumber \\
& & \left. + 16 \d_{\m \n} \Box H[ 1, \Box \; ; \; 1, 1] - 16 \d_{\m \n} \Box H[ 1, \pa_{\l} \; ; \; 1, \pa_{\l}] \; \right]
\er

We now have to proceed as in the QED and SQED examples. Expressions that have a d'Alembertian can be put in terms of $I^1$, and their renormalization is straightforward. The rest of the integrals can be found in the list of section \ref{Overlapping_div} (or can be easily expressed in terms of integrals of that list). Finally , we arrive to 
\br
< B_{\m}^a(x) B_{\n}^b(0) >_{k \; R} &=& \frac{g^4 C_A^2 \d^{ab}}{32 (4 \pi^2)^3} \left[ \pa_{\m} \pa_{\n} \Box \frac{- \frac{27}{4} \ln^2 x^2 M^2 - \frac{45}{4} \ln x^2 M^2}{x^2} + \right. \nonumber \\
& & \left. + \d_{\m \n} \Box \Box \frac{ \frac{27}{4} \ln^2 x^2 M^2 + \frac{33}{4} \ln x^2 M^2}{x^2} \right] + \ldots
\er

\subsubsection{Two loop final results}

Proceeding in the same way as for diagrams (a) and (k) (see appendix \ref{ap_two_loop}), we can obtain the renormalization of the rest of the diagrams, that we list here. 
\br
(a) ~~~ & & \pa_{\m} \pa_{\n} \Box \frac{ - \frac{1}{3} \ln^2 x^2 M^2 - \frac{8}{9} \ln x^2 M^2}{x^2} + \d_{\m \n} \Box \Box \frac{ \frac{1}{3} \ln^2 x^2 M^2 + \frac{11}{9} \ln x^2 M^2}{x^2} + \ldots \nonumber \\
(b) ~~~ & & \pa_{\m} \pa_{\n} \Box \frac{ -\frac{25}{3} \ln^2 x^2 M^2 - \frac{86}{9} \ln x^2 M^2}{x^2} + \d_{\m \n} \Box \Box \frac{ \frac{25}{3} \ln^2 x^2 M^2 + \frac{71}{9} \ln x^2 M^2}{x^2} + \ldots \nonumber \\
(c) ~~~ & & \d_{\m \n} \Box \Box \frac{ \frac{1}{2} \ln x^2 M^2}{x^2} + \ldots \nonumber \\
(d) ~~~ & & \d_{\m \n} \Box \Box \frac{ - \frac{9}{2} \ln x^2 M^2}{x^2} + \ldots \nonumber \\
(e) ~~~ & & ( \pa_{\m} \pa_{\n} - \d_{\m \n} \Box ) \Box \frac{12 \ln x^2 M^2}{x^2} + \ldots \nonumber \\
(f) ~~~ & & \pa_{\m} \pa_{\n} \Box \frac{ \frac{1}{3} \ln^2 x^2 M^2 - \frac{1}{9} \ln x^2 M^2}{x^2} + \d_{\m \n} \Box \Box \frac{- \frac{1}{3} \ln^2 x^2 M^2 - \frac{11}{9} \ln x^2 M^2}{x^2} + \ldots \nonumber \\
(g) ~~~ & & \pa_{\m} \pa_{\n} \Box \frac{ \frac{5}{12} \ln^2 x^2 M^2 + \frac{19}{36} \ln x^2 M^2}{x^2} + \d_{\m \n} \Box \Box \frac{- \frac{5}{12} \ln^2 x^2 M^2 - \frac{7}{36} \ln x^2 M^2}{x^2} + \ldots  \nonumber \\
(h) ~~~ & & \pa_{\m} \pa_{\n} \Box \frac{ \frac{9}{4} \ln^2 x^2 M^2 + \frac{21}{4} \ln x^2 M^2}{x^2} + \d_{\m \n} \Box \Box \frac{ - \frac{9}{4} \ln^2 x^2 M^2 + \frac{15}{4} \ln x^2 M^2}{x^2} + \ldots \nonumber \\
(i) ~~~ & & \pa_{\m} \pa_{\n} \Box \frac{-\frac{1}{12} \ln^2 x^2 M^2 - \frac{17}{36} \ln x^2 M^2}{x^2} + \d_{\m \n} \Box \Box \frac{ \frac{1}{12} \ln^2 x^2 M^2 + \frac{29}{36} \ln x^2 M^2}{x^2} + \ldots \nonumber \\
(j) ~~~ & & \pa_{\m} \pa_{\n} \Box \frac{ \frac{1}{2} \ln^2 x^2 M^2 + \frac{1}{2} \ln x^2 M^2}{x^2} + \d_{\m \n} \Box \Box \frac{- \frac{1}{2} \ln^2 x^2 M^2 - \frac{1}{2} \ln x^2 M^2}{x^2} + \ldots \nonumber \\
(k) ~~~ & & \pa_{\m} \pa_{\n} \Box \frac{ - \frac{27}{4} \ln^2 x^2 M^2 - \frac{45}{4} \ln x^2 M^2 }{x^2} + \d_{\m \n} \Box \Box \frac{ \frac{27}{4} \ln^2 x^2 M^2 + \frac{33}{4} \ln x^2 M^2 }{x^2} + \ldots \nonumber \\
\er

In this list all the expressions have a common factor of $\frac{1}{32(4 \pi^2)^3}g^4 C_A^2 \d^{ab}$.

Finally, adding up all the terms we obtain the divergent part of the two loop renormalized expression for the background field selfenergy as
\br
< B_{\m}^a (x) B_{\n}^b (0) >_R &=& - \frac{g^4 C_A^2 \d^{ab}}{2(4 \pi^2)^3} ( \pa_{\m} \pa_{\n} - \d_{\m \n} \Box ) \Box \frac{ \ln x^2 M^2}{x^2} + \ldots \label{2_loop}
\er

\subsection{Renormalization group equation}
\label{sec_RG}

With the previously obtained expressions for the one and two loop corrections of the background field propagator we can easily obtain the first two coefficients of the expansion of the beta function.

\subsubsection{RG equation for quantum fields}

We have to consider the renormalization group equation for the quantum field propagator, as with this we can find the value of $\g_{\xi}$, which is the function that takes care of the running of the gauge parameter. The only relevant coefficient of the expansion of $\g_{\xi}$ in our two loop background calculation will be the first one, so that we only need the one loop correction of the quantum gauge field propagator. 

First, we write the effective action as 
\br
\G &=& \frac{1}{2} \int d^4 x d^4 y \; A_{\m}^a (x) A_{\n}^b (y) \G^{(2) ab}_{\m \n}(x-y) + {\cal{O}} (A^3)
\er

If we consider the part of the Yang-Mills lagrangian which depends only in the quantum gauge fields, we can see that in a generic gauge is of the following form\br
\frac{1}{4} F_{\m \n}^a F_{\m \n}^a + \frac{1}{2} (1 + \xi) ( \pa_{\m} A_{\m}^a)(\pa_{\n} A_{\n}^a ) 
\er

With this, the free action plus the gauge fixing term becomes
\br
S_0 + S_{g.f.} &=& \frac{1}{2} \int d^4 x d^4 y \; A_{\m}^a \left( - \d_{\m \n} \Box \d (x) - \xi \pa_{\m} \pa_{\n} \d (x) \right) A_{\n}^a + {\cal{O}} (A^3)
\er

Using the one loop calculation (\ref{1loop_quantum}), $\G^{(2) ab}_{\m \n}$ can be written as
\br
\G^{(2)ab}_{\m \n} &=& - \d_{\m \n} \Box \d (x) - \xi \pa_{\m} \pa_{\n} \d (x) + \d^{ab} (\pa_{\m} \pa_{\n} - \d_{\m \n} \Box) \left[ \frac{5 g^2 C_A}{48 \pi^2 (4 \pi^2)} \Box \frac{\ln x^2 M^2}{x^2} + \right. \nonumber \\
&+& \left. \frac{g^2 C_A}{72 \pi^2 (4 \pi^2)} \d (x) \right] + {\cal{O}} (g^4)
\er

Inserting this into the RG equation
\br
\left[ M \frac{\pa}{\pa M} + \b (g) \frac{\pa}{\pa g} + \g_{\xi} \frac{\pa}{\pa \xi} - 2 \g_A \right] \G^{(2)ab}_{\m \n} |_{\xi=0} =0
\er

it is easy to obtain the desired result (along with the one loop coefficient for the anomalous dimension of gauge fields $\g_A$)
\br
\g_{\xi} &=& - \frac{5 C_A}{24 \pi^2} g^2 + \cdots \nonumber \\
\g_A &=& - \frac{5 C_A}{48 \pi^2} g^2 + \cdots \ \label{g_xi}
\er

\subsubsection{RG equation for background fields}
If we define 
\begin{equation}
\G^{(2) ab}_{\m \n} (x) = (\pa_{\m} \pa_{\n} - \d_{\m \n} \Box)\d^{ab} \G^{(2)} (x)\end{equation}

the equation we need to consider is
\br
\left[ M \frac{\pa}{\pa M} + \b (g) \frac{\pa}{\pa g} + \gamma_{\xi} \frac{\pa}{\pa \xi} - 2 \g_{B} \right] \G^{(2)}|_{\xi=0} =0 \label{YM_RG_eq}
\er

If the background field is redefined as $B^{\prime} = g B$ this implies $\gamma_{B} =0$ (due to the fact that in the background field approach the charge and background field renormalizations are related: $Z_{g} = Z_{B}^{-1/2}$ \cite{Abbott:1980hw}). Then, with the one loop contribution (\ref{1_loop}), the gauge fixing renormalization (\ref{gauge_fix_ren}) and the two loop contribution (\ref{2_loop}) the effective action for the background fields is
\br
\G^{(2)}(x) &=& \frac{1}{g^2} \d (x) + \frac{11 C_A}{48 \pi^2 (4 \pi^2)}\Box \frac{\ln x^2 M^2}{x^2} + \frac{ C_A}{72 \pi^2} \d (x) +  \frac{\xi C_A}{8 \pi^2} \d (x) + \nonumber \\
&+& \frac{g^2 C_A^2}{2 (4 \pi^2)^3} \Box \frac{ \ln x^2 M^2}{x^2} + \ldots \nonumber \\ \label{2loop_eff_action_YM}
\er

Using (\ref{2loop_eff_action_YM}) and the previously obtained expansion for $\g_{\xi}$ (\ref{g_xi}) into the RG equation (\ref{YM_RG_eq}), we can straightforwardly evaluate the first two coefficients of the expansion of the beta function to be
\br
\b (g) &=& \b_1 g^3 + \b_2 g^5 + {\cal{O}}(g^7) \nonumber \\
\b_1 &=& - \frac{11 C_A}{48 \pi^2} \nonumber \\
\b_2 &=& - \frac{17 C^2_A}{24 (4 \pi^2)^2}
\er

These results agree with those previously obtained in the literature \cite{Abbott:1980hw}.

\section{Conclusions}

In this paper we have shown how the results of CDR, found for the one loop case, can be very useful in two loop calculations. In those cases, the use of one loop CDR allows us to fix all the ambiguities that appear at one loop level, which implies that the coefficients of the divergent parts of the total expression are fixed {\em{a priori}}. No Ward identity is needed to be used to relate different scales. The procedure distinguishes between the cases of diagrams with nested divergences (where we renormalize from inside to outside the diagram, applying first CDR and after that normal DR) and diagrams with overlapping divergences. In order to deal with the latter, we have found a list of renormalized integrals that can be applied to various two loop two point function calculations. 

As an example of all this, we have found applying our procedure the two loop beta function coefficient of QED, SuperQED and Yang-Mills theory. 

\section*{Acknowledgments}

I would like to thank J. Mas for useful discussions and critical reading of the manuscript. Also I'm very grateful to M. Perez-Victoria and J.I. Latorre for sharing their insight with me on CDR and the two loop differential renormalization of QED, respectively. Finally, I would like to thank M.Gomez-Reino for her critical reading of some parts of the manuscript.

\appendix

\section{Integrals with overlapping divergences}
\label{ap_integrales}

Here we present the explicit calculation of the list of integrals of section \ref{Overlapping_div}. As there, we define the final results in terms of a variable $z=x-y$ and denote $\pa_{\m} \equiv \pa_{\m}^x$. Also, in each final expression $\ldots$ stands for the finite contribution that we are not taking into account.

\begin{itemize}

\item $H[1,1 \; ; \; 1,1] \\
H[\pa_{\m},1 \; ; \; 1,1] \\
H[ 1, \pa_{\l} \; ; \; 1, \pa_{\l}]$

These integrals are obtained by means of Gegenbauer Polynomials \cite{Freedman:1991tk,Song}.


\item $ \pa_{\l}^x H[ 1, \pa_{\m} \; ; \; 1, \pa_{\l}]$

Contracting (\ref{rel_int1}) with $\d_{\n \l}$ we find
\br
\pa_{\l}^x H[ 1, \pa_{\m} \; ; \; 1, \pa_{\l}] &=& \frac{1}{2} \pa_{\m}^x H [ 1, 1 \; ; \; \Box, 1] - H[ 1, \pa_{\m} \; ; \; 1, \Box ] - \frac{1}{2} \Box H [ 1, \pa_{\m} \; ; \; 1,1] \nonumber \\
&=& \frac{1}{2} \pa_{\m} ( \D I^1 ) - ( \D \pa_{\m} I^1 ) + \frac{a}{4} ( \pa_{\m} \d ) \nonumber \\
&\stackrel{R}{\rightarrow}& - \frac{1}{32 (4 \pi^2)^3} \pa_{\m} \Box \frac{ \frac{1}{2} \ln z^2 M^2}{z^2} + \ldots
\er

\item $H[ \pa_{\m} \pa_{\l}, \pa_{\l} \; ; \; 1,1]$

In this case, no integral relation is used
\br
H[ \pa_{\m} \pa_{\l}, \pa_{\l} \; ; \; 1,1] &=& \frac{1}{2} \pa_{\m}^x H[ \pa_{\l}, \pa_{\l} \; ; \; 1,1] \nonumber \\
&=& \frac{1}{2} \pa_{\m}^x \pa_{\l}^x H[1, \pa_{\l} \; ; \; 1, 1] - \frac{1}{2} \pa_{\m}^x H[1, \Box \; ; \; 1,1] \nonumber \\
&=& \frac{1}{2} \pa_{\m} ( \D I^1 ) -  \frac{a}{4} ( \pa_{\m} \d) \nonumber \\
&\stackrel{R}{\rightarrow}&  \frac{1}{32(4 \pi^2)^3} \pa_{\m} \Box \frac{ - \frac{1}{2} \ln^2 z^2 M^2 - \ln z^2 M^2}{z^2} + \ldots
\er

\item $ \pa_{\l}^x H[ 1, \pa_{\m} \; ; \; \pa_{\n} \pa_{\l}, 1] \\
\pa_{\l}^x H[ 1, \pa_{\l} \; ; \; \pa_{\m} \pa_{\n}, 1] \\
H[ 1, \pa_{\l} \; ; \; \pa_{\l} \pa_{\m},1]$

First of all, the third integral (\ref{int6}) will be evaluated with relation (\ref{rel_int1})
\br
H[ 1, \pa_{\l} \; ; \; \pa_{\l} \pa_{\n},1] &=& \frac{1}{2} \pa_{\l}^x H [1, \pa_{\l} \; ; \; 1, \pa_{\n} ] + \frac{1}{2} \pa_{\l}^x H[ 1, 1 \; ; \; 1, \pa_{\l} \pa_{\n} ] + \frac{1}{2} \pa_{\n}^x H[ 1, \pa_{\l} \; ; \; 1, \pa_{\l}] + \nonumber \\
&+& \frac{1}{2} \pa_{\n}^x \pa_{\l}^x H[ 1, \pa_{\l} \; ; \; 1,1]
\er

Using the previous results
\br
H^R[ 1, \pa_{\l} \; ; \; \pa_{\l} \pa_{\n},1] &=& \frac{1}{32 (4 \pi^2)^3} \pa_{\m} \Box \frac{\frac{1}{8} \ln^2 z^2 M^2 - \frac{7}{8} \ln z^2 M^2 }{z^2} + \ldots
\er

But this integral along with the other two, can be obtained with other method. We can apply the CDR decomposition into trace part, traceless part and local term (\ref{CDR_T}) to the divergent subdiagram  $ (\pa_{\m}^x \pa_{\n}^x \D_{yu}) \D_{yv} \D_{uv}$. Let us consider  this in the general integral 
\br
\int &d^4 u d^4 v& \D_{xu} ( \pa_{\rho}^x \D_{xv} ) ( \pa_{\eps}^y \pa_{\s}^y \D_{yu} ) \D_{yv} \D_{uv} = \nonumber \\
&-& \frac{1}{4} \d_{\eps \s} ( \D \pa_{\rho} I^1 )_R - \frac{\d_{\eps \rho}}{256 \pi^2} \pa_{\rho} \D^2_R - \frac{16}{(4 \pi^2)^5} I_{\rho \eps \s \; R} \label{I_rho_eps_sig} 
\er

where $I_{\rho \eps \s}$ stands for the traceless part. The one loop ambiguity fixed by CDR is reflected in the second term of (\ref{I_rho_eps_sig}) that at two loops has becamed a logarithm of the scale. In the renormalization of the traceless part normal differential renormalization will be used, leaving ambiguities (local terms) not fixed.

The expression for $I_{\rho \eps \s}$ is 
\br
I_{ \rho \eps \s \; R } &=& B \frac{ x_{\eps} x_{\s} x_{\rho}}{x^8} - \frac{1}{2} A \frac{x_{\rho}}{x^6} \d_{\eps \s} + ( A - \frac{1}{2} B) \left[ \frac{x_{\eps}}{x^6} \d_{\rho \s} + \frac{x_{\s}}{x^6} \d_{\rho \eps} \right] |_R
\er

or in the form of the integrals being discussed
\br
I_{\l \l \m \; R} &=& - \frac{3}{8} ( 4 \pi^2)^2 ( 3A - B ) \pa_{\m} \D^2_R \label{I_lam_lam_mu}\\
\pa_{\l} I_{\m \l \n \; R} &=& - (4 \pi^2)^2 ( 3A - B ) \left[ \frac{1}{24} \pa_{\m} \pa_{\n} \D^2_R + \frac{2}{3} ( 4 \pi^2) \d_{\m \n} \D^3_R \right] \\
\pa_{\l} I_{\l \m \n \; R} &=& (4 \pi^2)^2 (3A - B) \left[ - \frac{1}{6} \pa_{\m} \pa_{\n} \D^2_R + \frac{1}{3} (4 \pi^2) \d_{\m \n} \D^3_R \right]
\er

The value of $(3A-B)$ is easily obtained using (\ref{I_lam_lam_mu}), because this corresponds to integral (\ref{int6}) that was obtained previously, 

I.e.
\br
\int &d^4 u d^4 v& \D_{xu} ( \pa_{\l}^x \D_{xv}) ( \pa_{\l}^y \pa_{\n}^y \D_{yu}) \D_{yv} \D_{uv} = \nonumber \\
&=& \frac{1}{32 (4 \pi^2)^3} \pa_{\m} \Box \frac{\frac{1}{8} \ln^2 z^2 M^2 - \frac{7}{8} \ln z^2 M^2 }{z^2} \nonumber \\
&=& \frac{1}{32 (4 \pi^2)^3} \pa_{\n} \Box \frac{ \frac{1}{8} \ln^2 z^2 M^2 + \frac{1}{4} \ln z^2 M^2}{z^2}- \frac{16}{(4 \pi^2)^5} I_{\rho \eps \s \; R}
\er 

which implies that
\br
(3A - B) &=& \frac{3 \pi^4}{8}
\er

With this result the evaluation of (\ref{int8}) and (\ref{int10}) are straightforward
\br
\pa_{\l}^x H^R[ 1, \pa_{\m} \; ; \; \pa_{\n} \pa_{\l}, 1] &=& - \frac{1}{4} \pa_{\n} ( \D \pa_{\m} I^1 )_R - \frac{1}{256 \pi^2} \pa_{\m} \pa_{\n} \D^2_R - \frac{16}{(4 \pi^2)^5} \pa_{\l} I_{\m \l \n \; R} \nonumber \\
&=& \frac{1}{32(4 \pi^2)^3} \left[ \pa_{\m} \pa_{\n} \Box \frac{ \frac{1}{8} \ln^2 z^2 M^2 + \frac{1}{8} \ln z^2 M^2}{z^2} + \d_{\m \n} \Box \Box \frac{-\frac{1}{4} \ln z^2 M^2}{z^2} \right] + \ldots \nonumber \\
\er

\br
\pa_{\l}^x H^R[ 1, \pa_{\l} \; ; \; \pa_{\m} \pa_{\n}, 1] &=& - \frac{1}{4} \d_{\m \n} \pa_{\l} ( \D \pa_{\l} I^1 )_R - \frac{\d_{\m \n}}{256 \pi^2} \Box \D^2_R - \frac{16}{(4 \pi^2)^5} \pa_{\l} I_{\l \m \n \; R} \nonumber \\
&=& \frac{1}{32 (4 \pi^2)^3} \left[ \pa_{\m} \pa_{\n} \Box \frac{ - \frac{1}{2} \ln z^2 M^2}{z^2} + \d_{\m \n} \Box \Box \frac{\frac{1}{8} \ln^2 z^2 M^2 + \frac{3}{8} \ln z^2 M^2}{z^2} \right] + \ldots \nonumber \\
\er

\item $\pa_{\l}^x H[1, \pa_{\l} \; ; \; 1, \pa_{\m} \pa_{\n}]$

Using integral relation \ref{rel_int1}
\br
\pa_{\l}^x H[1,\pa_{\l} \; ; \; 1, \pa_{\m} \pa_{\n}] &=& \pa_{\l}^x H[ 1, \pa_{\l} \; ; \; \pa_{\m} \pa_{\n}, 1] - \pa_{\l}^x \pa_{\m}^x H [1, \pa_{\l} \; ; \; 1, \pa_{\n}] - \pa_{\n}^x \pa_{\l}^x H[ 1, \pa_{\l} \; ; \; 1,\pa_{\m}] - \nonumber \\
&-& \pa_{\n} \Box H[1, \pa_{\m} \; ; \; 1,1] \nonumber \\
&\stackrel{R}{\rightarrow}&  \frac{1}{32 (4 \pi^2)^3} \left[ \pa_{\m} \pa_{\n} \Box \frac{ \frac{1}{2} \ln z^2 M^2}{z^2} + \d_{\m \n} \Box \Box \frac{\frac{1}{8} \ln^2 z^2 M^2 + \frac{3}{8} \ln z^2 M^2}{z^2} \right] + \ldots \nonumber \\
\er

\item $H[1, \pa_{\m} \; ; \; 1, \pa_{\n}]$

Considering (\ref{int8}) and applying (\ref{rel_int1})
\br
\pa_{\l}^x H[ 1, \pa_{\m} \; ; \; \pa_{\n} \pa_{\l} , 1] &=& \frac{1}{2} \Box H [1, \pa_{\m} \; ; \; 1, \pa_{\n}] + \frac{1}{2} \pa_{\m}^x \pa_{\l}^x H[ 1, 1 \; ; \; 1, \pa_{\n} \pa_{\l}] + \nonumber \\
&+& \frac{1}{2} \pa_{\n}^x \pa_{\l}^x H[ 1, \pa_{\m} \; ; \; 1, \pa_{\l}] + \frac{1}{2} \pa_{\n}^x \Box H[ 1, \pa_{\m} \; ; \; 1,1]
\er

Remembering previous results
\br
\pa_{\l}^x H^R [1, \pa_{\m} \; ; \; \pa_{\n} \pa_{\l},1] &=& \frac{1}{2} \Box H^R [1, \pa_{\m} \; ; \; 1, \pa_{\n}] + \frac{1}{32(4 \pi^2)^3} \pa_{\m} \pa_{\n} \Box \frac{ \frac{1}{8} \ln^2 z^2 M^2 + \frac{1}{8} \ln z^2 M^2}{z^2} + \ldots \nonumber \\
\er

So that
\br
\Box H^R[1, \pa_{\m}\; ; \; 1, \pa_{\n}] &=& \frac{1}{32 (4 \pi^2)^3} \d_{\m \n} \Box \Box \frac{- \frac{1}{2} \ln z^2 M^2}{z^2} + \ldots \nonumber \\
\er

\item $H[1,1 \; ; \; \pa_{\m} \pa_{\n}, 1]$ 

Using  (\ref{int10}), (\ref{int10a}) and the identity
\br
\Box H[ 1,1 \; ; \; \pa_{\m} \pa_{\n},1] &=& \pa_{\l}^x H[ 1, \pa_{\l} \; ; \; \pa_{\m} \pa_{\n},1] + \pa_{\l}^x H[ \pa_{\l},1 \; ; \; \pa_{\m} \pa_{\n},1]
\er

we can easily arrive to
\br
\Box H^R[1,1 \; ; \; \pa_{\m} \pa_{\n}, 1] &=& \frac{1}{32(4 \pi^2)^3} \d_{\m \n} \Box \Box \frac{ \frac{1}{4} \ln^2 z^2 M^2 + \frac{3}{4} \ln z^2 M^2}{z^2} + \ldots
\er

\item $\pa_{\l}^x H[1,1 \; ; \; \pa_{\l} \pa_{\n} , \pa_{\m}]$

Using (\ref{rel_int3}) and (\ref{int11}) 
\br
\pa_{\l}^x H[1,1 \; ; \; \pa_{\l} \pa_{\n} , \pa_{\m}] &=& \frac{1}{2} \Box H[ 1,1 \; ; \; \pa_{\m} \pa_{\n} ,1] - \frac{1}{4} \pa_{\m}^y \pa_{\n}^y \pa_{\l}^y H[ 1,1 \; ; \; \pa_{\l},1] - \frac{1}{4} \pa_{\n}^y \Box H[1,1 \; ; \; \pa_{\m} , 1] \nonumber \\
&\stackrel{R}{\rightarrow}& \frac{1}{32 (4 \pi^2)^3} \d_{\m \n} \Box \Box \frac{ \frac{1}{8} \ln^2 z^2 M^2 + \frac{3}{8} \ln z^2 M^2}{z^2} + \ldots
\er

\item $\pa_{\l}^x H[ 1,1 \; ; \; \pa_{\m} \pa_{\n} , \pa_{\l}]$

With (\ref{rel_int3}) and (\ref{int5})
\br
\pa_{\l}^x H[ 1,1 \; ; \; \pa_{\m} \pa_{\n} , \pa_{\l}] &=& \frac{1}{4} \pa_{\n}^x \Box H[ 1,1 \; ; \; \pa_{\m},1] + \frac{1}{4} \pa_{\l}^x \pa_{\m}^x \pa_{\n}^x H[ 1,1 \; ; \; \pa_{\l},1] + \frac{1}{2} \pa_{\l}^x \pa_{\m}^x H[ 1,1 \; ; \; \pa_{\l} \pa_{\n},1] \nonumber \\
&\stackrel{R}{\rightarrow}& \frac{1}{32 (4 \pi^2)^3} \pa_{\m} \pa_{\n} \Box \frac{\frac{1}{8} \ln^2 z^2 M^2 + \frac{3}{8} \ln z^2 M^2}{z^2} + \ldots
\er

\item $H[1, \pa_{\m} \pa_{\l} \; ; \; \pa_{\n} \pa_{\l}, 1]$

In this case the CDR decomposition into trace part, traceless part and fixed local term (\ref{CDR_T})  will be used again, as in (\ref{int8}) and (\ref{int10})
\br
H[1, \pa_{\m} \pa_{\l} \; ; \; \pa_{\n} \pa_{\l}, 1] &=& - \frac{1}{4} ( \D \pa_{\m} \pa_{\n} I^1 )_R - \frac{1}{4} [ \D ( \pa_{\m} \pa_{\n} - \frac{1}{4} \d_{\m \n} \Box ) I^1 ]_R - \nonumber \\
&-& \frac{1}{128 \pi^2} (\D \pa_{\m} \pa_{\n} \D)_R - \frac{1}{128 \pi^2} [ \D ( \pa_{\m} \pa_{\n} - \frac{1}{4} \d_{\m \n} \Box ) \D ]_R + \nonumber \\
&+& \frac{64}{(4 \pi^2)^5} I_{\m \l \n \l \; R}
\er

where $I_{\m \l \n \l}$ stands for the integral with the traceless parts. This was calculated in \cite{Haagensen:1992vz}, and the result was found there to be
\br
\frac{64}{(4 \pi^2)^5} I_{\m \l \n \l \; R} &=& \frac{5}{96(4 \pi^2)} \pa_{\m} \pa_{\n} \D^2_R + \frac{13}{48} \d_{\m \n} \D^3_R
\er

Adding up all the terms, it is easy to arrive to
\br
H^R[1, \pa_{\m} \pa_{\l} \; ; \; \pa_{\n} \pa_{\l}, 1] &=& \frac{1}{32 (4 \pi^2)^3} \left[ \pa_{\m} \pa_{\n} \Box \frac{ \frac{1}{6} \ln^2 z^2 M^2 - \frac{5}{36} \ln z^2 M^2}{z^2} + \right. \nonumber \\
&+& \left. \d_{\m \n} \Box \Box \frac{ - \frac{1}{24} \ln^2 z^2 M^2 - \frac{29}{72} \ln z^2 M^2}{z^2} \right] + \ldots \nonumber \\
\er

\item $H[1, \pa_{\m} \pa_{\l} \; ; \; 1 , \pa_{\l} \pa_{\n} ]$

In this case, applying (\ref{rel_int1}), (\ref{int5}), (\ref{int6}), (\ref{int8}) and (\ref{int14}) we get
\br
H[1, \pa_{\m} \pa_{\l} \; ; \; 1 , \pa_{\l} \pa_{\n} ] &=& H[ 1, \pa_{\m} \pa_{\l} \; ; \; \pa_{\n} \pa_{\l},1] - \pa_{\l}^x H[ 1, \pa_{\m} \pa_{\l} \; ; \; 1, \pa_{\n}] - \pa_{\n}^x H[1, \pa_{\m} \pa_{\l} \; ; \; 1, \pa_{\l}]- \nonumber \\&-& \pa_{\n}^x \pa_{\l}^x H[ 1, \pa_{\m} \pa_{\l} \; ; \; 1,1] \nonumber \\
&\stackrel{R}{\rightarrow}& \frac{1}{32 (4 \pi^2)^3} \left[ \pa_{\m} \pa_{\n} \Box \frac{\frac{1}{6} \ln^2 z^2 M^2 + \frac{49}{36} \ln z^2 M^2}{z^2} +\right. \nonumber \\
&+& \left.\d_{\m \n} \Box \Box \frac{- \frac{1}{24} \ln^2 z^2 M^2 - \frac{11}{72} \ln z^2 M^2}{z^2} \right] + \ldots
\er

\end{itemize}

\section{Expansion of $\Pi_{\m \n}^{(2 \; b)}$}
\label{ap_pi_mu_nu}
The complete expanded expression of the two loop diagram b of QED is
\br
\Pi^{(2 \; b)}_{\m \n} = e^4 &\left[\right.& - 8 \d_{\m \n} \Box H[ 1 , \pa_{\l} \; ; \; \pa_{\l}, 1] + 16 \pa_{\m}^x H[ 1, \pa_{\n} \; ; \; \Box, 1] - 8 \d_{\m \n} \pa_{\l}^x H[ 1, \pa_{\l} \; ; \; \Box , 1] - \nonumber \\
& & - 16 \pa_{\m}^x H[ 1, \pa_{\l} \; ; \; \pa_{\l} \pa_{\n}, 1] + 16 \pa_{\l}^x H [ 1, \pa_{\l} \; ; \; \pa_{\m} \pa_{\n} ,1] - 16 \pa_{\l}^x H [ 1, \pa_{\m} \; ; \; \pa_{\l} \pa_{\n},1] - \nonumber \\
& & - 16 \pa_{\m} H [ 1, \Box \; ; \; \pa_{\n}, 1] + 8 \d_{\m \n} \pa_{\l}^x H[ 1, \Box \; ; \; \pa_{\l}, 1] + 16 \pa_{\l}^x H[ 1, \pa_{\l} \pa_{\m} \; ; \; \pa_{\n},1] - \nonumber \\
& & - 16 \pa_{\l}^x H[ 1, \pa_{\m} \pa_{\n} \; ; \; \pa_{\l}, 1] + 16 \pa_{\n}^x H[ 1, \pa_{\l} \pa_{\m} \; ; \; \pa_{\l}, 1] - 16 H[ 1, \Box \; ; \; \pa_{\m} \pa_{\n},1] + \nonumber \\
& & \left. + 8 \d_{\m \n} H[ 1, \Box \; ; \; \Box ,1 ] + 32 H[ 1 , \pa_{\m} \pa_{\l} \; ; \; \pa_{\n} \pa_{\l}, 1] - 16 H[ 1 , \pa_{\m} \pa_{\n} \; ; \; \Box , 1] \; \right]
\er

\section{Two loop diagrams of Yang-Mills theory}
\label{ap_two_loop}
We remind that in these diagrams $D_{\m}$ denotes a derivative acting over one of the external fields, and again $\ldots$ stand for the two loop finite contributions that we are not taking into account.

\subsection{Diagram (b)}

\subsubsection{Definitions}

When obtaining this diagram some quantities are defined in order to simplify the expressions (as we defined $I^1$). They are
\br
I^0 (x-y) &=& \int d^4 u d^4 v \; \D_{xu} \D_{yv} \D^2_{uv} \\
I^0_{\m} (x-y) &=& \int d^4 u d^4 v \; \D_{xu} \D_{yv} ( \D_{uv} \pa_{\m}^u  \D_{uv} ) \\
I_{\m \n}^0 (x-y) &=& \int d^4 u d^4 v \; \D_{xu} \D_{yv} ( \D_{uv} \pa_{\m}^u  \pa_{\n}^u  \D_{uv} )
\er

We will detail the renormalization of the terms formed with these integrals that appear in diagram (b): $ \D \Box I^0$, $ \D \pa_{\m} \pa_{\n} I^0$ and $\D I^0_{\m \n}$.

\begin{itemize}

\item $I^{0}$ has UV and IR divergences. This expression was studied in \cite{Mas:2002xh} and evaluated as
\br
I^0_R &=& \frac{1}{32 (4 \pi^2)^2} \left[ \ln^2 x^2 M^2_{IR} + 2 \ln x^2 M^2_{IR} ( 1 - \ln x^2 M^2) + b_{IR} \right]
\er

It is clear also that $\Box I^0 = - I^1$, and it is not difficult to prove that
\br
(\D \pa_{\m} \pa_{\n} I^0 )_R &=& \frac{1}{32(4 \pi^2)^3} \left[ \pa_{\m} \pa_{\n} \frac{ \ln x^2 M^2}{x^2} + \d_{\m \n} \Box \frac{ \frac{1}{4} \ln^2 x^2 M^2 + \frac{1}{4} \ln x^2 M^2}{x^2} \right] + \dots
\er

\item The renormalization of $I^0_{\m}$ is straightforward, once we recall that CDR imposes $I^0_{\m \; R} = \frac{1}{2} \pa_{\m}^x I^0_R$

\item With $I_{\m \n}^0$, applying CDR to the subdivergency we find
\br
I_{\m \n}^0 &=& \frac{1}{3} \int d^4 u d^4 v \; \D_{xu} \D_{yv} ( \pa_{\m} \pa_{\n} - \frac{1}{4} \d_{\m \n} \Box ) (\D^2_{uv})_{R} + \nonumber \\
&+& \frac{1}{288 \pi^2} \int d^4 u d^4 v \; \D_{xu} \D_{yv} ( \pa_{\m} \pa_{\n} - \d_{\m \n} \Box) \d (u-v) \nonumber \\
&=& \frac{1}{3} \pa_{\m} \pa_{\n} I^0_R - \frac{1}{12} \d_{\m \n} \Box I^0_R + \frac{1}{72 (4\pi^2)} \pa_{\m} \pa_{\n} \int d^4 u \; \D_{xu} \D_{yu} + \frac{\d_{\m \n}}{72 (4\pi^2)} \D \nonumber 
\er

or, which is the same
\br
(\D I_{\m \n}^0)_R &=& \frac{1}{32(4 \pi^2)^3} \left[ \pa_{\m} \pa_{\n} \frac{ \frac{1}{3} \ln x^2 M^2}{x^2} + \d_{\m \n} \Box \frac{ - \frac{1}{6} \ln x^2 M^2}{x^2} \right] + \ldots
\er

\end{itemize}

\subsubsection{Evaluation of diagram (b)}

This diagram is of the form
\br
<B_{\m}^a (x) B_{\n}^b (y) >_b &=& g^2 f^{ace} f^{bed} \int d^4 u d^4 v \; \D_{xu} \left[ - 2 \d_{\m \l} D_{\rho}^x + \d_{\l \rho} ( \stackrel{\leftarrow}{\pa_{\m}^x} - \pa_{\m}^x) + 2 \d_{\m \rho} D_{\l}^x \right] \times \nonumber \\
& & \times \D^{cd}_{\rho \s} (u-v) \D_{vy} \left[ - 2 \d_{\n \s} D_{\l}^y + \d_{\s \l} ( \pa_{\n}^y - \stackrel{\leftarrow}{\pa_{\n}^y }) + 2 \d_{\n \l} D_{\s}^y \right] \D_{xy} 
\er

Where $\D_{\m \n}$ is the one loop correction of the quantum gauge field propagator, which we have found to have the following non renormalized expression
\br
\D_{\m \n}^{ab} &=& g^2 C_A^2 \d^{ab} \left[ \pa_{\m} \pa_{\n} \D^2 - \d_{\m \n} \Box \D^2 + 4 \pa_{\m} ( \D \pa_{\n} \D) - 2 \d_{\m \n} \pa^{\l} ( \D \pa_{\l} \D ) - \right. \nonumber \\
&-& \left. 4 \D \pa_{\m} \pa_{\n} \D - \d_{\m \n} \D ( \Box \D ) \right]  
\er

It has to be noted that, in contrast with dimensional regularization, the renormalized one loop expression can not be used here. The reason is that the indices of the one loop insertion the will be contracted in a future step, and one of the rules of CDR is to make first all the index contractions before performing the renormalization. So that, we are only allowed to insert the bare one loop propagator. Then, the diagram can be put as 
\br
<B_{\m}^a (x) B_{\n}^b (y) >_b =  - g^2 C_A \d^{ab} \int & d^4 u d^4 v&  - 4 \pa_{\m}^x \pa_{\rho}^x \left( \D_{xu} \D_{\rho \n} (u-v) \D_{vy} \D_{xy} \right) \nonumber \\
& & + 4 \d_{\m \n} \pa_{\rho}^x \pa_{\s}^x \left( \D_{xu} \D_{\rho \s} (u-v) \D_{vy} \D_{xy} \right) \nonumber \\
& & + \D_{xu} ( \stackrel{\leftarrow}{\pa_{\m}^x} - \pa_{\m}^x ) \D_{\rho \rho} (u-v) \D_{vy} ( \pa_{\n}^y - \stackrel{\leftarrow}{\pa_{\n}^y} ) \D_{xy} \nonumber \\
& & + 4 \Box \left( \D_{xu} \D_{\m \n} (u-v) \D_{vy} \D_{xy} \right) \nonumber \\
& & - 4 \pa_{\n}^x \pa_{\s}^x \left( \D_{xu} \D_{\m \s} (u-v) \D_{vy} \D_{xy}  \right). \nonumber 
\er

Straightforwardly we write this expression in terms of the previously defined integrals as 
\br
<B_{\m}^a (x) B_{\n}^b (0) >_{b \; R} &=& g^4 C_A^2 \d^{b a} \left[ - 24 \pa_{\m} \pa_{\s} ( \D \pa_{\n} \pa_{\s} I^0) + 11 \pa_{\m} \pa_{\n} ( \D \Box I^0) + 32 \pa_{\m} \pa_{\s} ( \D I^0_{\n \s}) + \right. \nonumber \\
&+& 12 \d_{\m \n} \pa_{\s} \pa_{\rho} ( \D \pa_{\rho} \pa_{\s} I^0) - 16 \d_{\m \n} \Box ( \D \Box I^0) - 16 \d_{\m \n} \pa_{\rho} \pa_{\s} ( \D I^0_{\rho \s} ) +  \nonumber \\
&+& \left. 20 \pa_{\m} ( \D \pa_{\n} \Box I^0) - 20 \D \pa_{\m} \pa_{\n} \Box I^0 + 12 \Box ( \D \pa_{\m} \pa_{\n} I^0) - 16 \Box (\D I_{\m \n}^0) \right]_R \nonumber 
\er

And with the renormalized forms of $I^0$ and $I^0_{\m \n}$ is easy to arrive to the final renormalized expression
\br
<B_{\m}^a (x) B_{\n}^b (0) >_{b \; R} &=& \frac{g^4 C_A^2 \d^{a b}}{32 (4 \pi^2)^3} \left[ \pa_{\m} \pa_{\n} \Box \frac{- \frac{25}{3} \ln^2 x^2 M^2 - \frac{86}{9} \ln x^2 M^2}{x^2} \right. \nonumber \\
&+& \left.  \d_{\m \n} \Box \Box \frac{ \frac{25}{3} \ln^2 x^2 M^2 + \frac{71}{9} \ln x^2 M^2}{x^2} \right] + \ldots 
\er

\subsection{Diagram (c)}

The expression for this diagram is
\br
<B_{\m}^a (x) B_{\n}^b (0) >_{c \; R} &=& -  g^4 f^{acx} f^{xed} f^{bdy} f^{yec} \d_{\m \n} \D^3  \nonumber \\
&=& - \frac{1}{2} g^4 C_A^2 \d^{ab} \d_{\m \n} \D^3  \nonumber \\
&=& \frac{g^4 C_A^2 \d^{a b}}{32 (4 \pi^2)^3} \d_{\m \n} \Box \Box \frac{ \frac{1}{2} \ln x^2 M^2}{x^2} + \ldots 
\er

\subsection{Diagram (d)}

This diagram is similar to the previous one, and its result is
\br
<B_{\m}^a (x) B_{\n}^b (0) >_{d \; R} &=& \frac{9}{2} g^4 C_A^2 \d^{ab} \d_{\m \n} \D^3 \nonumber \\
&=& \frac{g^4 C_A^2 \d^{a b}}{32 (4 \pi^2)^3} \d_{\m \n} \Box \Box \frac{ -\frac{9}{2} \ln x^2 M^2}{x^2} + \ldots
\er

\subsection{Diagram (e)}

The bare expression for this diagram is
\br
<B_{\m}^a (x) B_{\n}^b (y) >_e &=& - \frac{1}{4} g^4 f^{acd} f^{bge} \int d^4 u \;  \D_{xu} \left[ - 2 \d_{\m \s} D_{\rho}^x + \d_{\rho \s} ( \stackrel{\leftarrow}{\pa_{\m}^x} - \pa_{\m}^x ) + 2 \d_{\m \rho} D^x_{\s} \right] \times \nonumber \\
& & \times  \D_{xu} \left[ f^{cex} f^{xgd} ( \d_{\rho \l} \d_{\eps  \s } - \d_{\rho \s} \d_{\eps \l} ) + f^{cgx} f^{xde} ( \d_{\rho \s} \d_{\eps \l} - \d_{\rho \eps} \d_{\s \l} ) + \right. \nonumber \\
& & +  \left. f^{cdx} f^{xeg} (\d_{\rho \eps} \d_{\l \s} - \d_{\rho \l} \d_{\eps \s} ) \right] \D_{yu} \left[ - 2 \d_{\n \eps} D_{\l}^y + \d_{\eps \l} ( \pa_{\n}^y - \stackrel{\leftarrow}{\pa_{\n}^y}) + \right. \nonumber \\
& & + \left. 2 \d_{\n \l} D_{\eps}^y \right] \D_{yu} \nonumber \\
&=& - 6 g^4 C_A^2 \d^{ab} ( \pa_{\m}^x \pa_{\n}^x - \d_{\m \n} \Box ) \int d^4 u \; \D^2_{xu} \D^2_{yu} \nonumber  
\er

The renormalized expression of the integral is easily obtained as
\br
\int d^4 u \; \frac{1}{(x-u)^4} \frac{1}{u^4} \rightarrow - \frac{\pi^2}{4} \Box \frac{\ln^2 x^2 M^2}{x^2} \nonumber 
\er

So that
\br
<B_{\m}^a (x) B_{\n}^b (0) >_{e \; R} &=& \frac{3}{8 (4 \pi^2)^3} g^4 C_A^2 \d^{ab} (\pa_{\m} \pa_{\n} - \d_{\m \n} \Box ) \Box \frac{ \ln^2 x^2 M^2}{x^2} + \ldots 
\er

\subsection{Diagram (f)}

This diagram is
\br
<B_{\m}^a (x) B_{\n}^b (y) >_f &=& 2 g^4 f^{acx} f^{xdf} f^{dce} f^{bef} \int d^4 u \; \D^2_{xu} ( \pa_{\m}^u \D_{uy} )( \pa_{\n}^y - \stackrel{\leftarrow}{\pa_{\n}^y} ) \D_{xy} + \nonumber \\
&+& 2 g^4 f^{afc} f^{ecd} f^{bfx} f^{xed} \int d^4 u \; \D_{xu} ( \stackrel{\leftarrow}{\pa_{\m}^x} - \pa_{\m}^x ) \D_{xy} ( \pa_{\n}^u \D_{yu} ) \D_{uy} \;. \nonumber
\er

Operating
\br
<B_{\m}^a (x) B_{\n}^b (0) >_{f \; R} &=& - g^4 C_A^2 \d^{ab} \left[ - 2 \pa_{\m} ( \D \pa_{\n} I^1 ) + 4 \D \pa_{\m} \pa_{\n} I^1 \right]_R \nonumber \\
&=& \frac{g^4 C_A^2 \d^{a b}}{32 (4 \pi^2)^3} \left[ \pa_{\m} \pa_{\n} \Box \frac{ \frac{1}{3} \ln^2 x^2 M^2 - \frac{1}{9} \ln x^2 M^2}{x^2} + \right. \nonumber \\
&+& \left. \d_{\m \n} \Box \Box \frac{ - \frac{1}{3} \ln^2 x^2 M^2 - \frac{11}{9} \ln x^2 M^2}{x^2} \right] + \ldots
\er

\subsection{Diagram (g)}
\br
<B_{\m}^a (x) B_{\n}^b (y) >_g &=& - 2 g^4 f^{acx} f^{xfd} f^{ecd} f^{bfe} \int d^4 u \; \d_{\m \s} \D_{xu} ( \pa_{\l}^u \D_{xu} ) \D_{uy} \left[ - 2 \d_{\n \l} D_{\s}^y + \right. \nonumber \\
&+& \left. \d_{\l \s} ( \pa_{\n}^y - \stackrel{\leftarrow}{\pa_{\n}^y} ) + 2 \d_{\n \s} D_{\l}^y \right] \D_{xy} \nonumber 
\er

It is easy to arrive to
\br
<B_{\m}^a (x) B_{\n}^b(0) >_{g \; R} &=& - g^4 C_A^2 \d^{ab} \left[ \frac{3}{2} \pa_{\m} ( \D \pa_{\n} I^1) - \D \pa_{\m} \pa_{\n} I^1 - \d_{\m \n} \pa_{\l} ( \D \pa_{\l} I^1)  \right]_R \nonumber \\
&=& \frac{g^4 C_A^2 \d^{ab}}{32 (4 \pi^2)^3} \left[ \pa_{\m} \pa_{\n} \Box \frac{ \frac{5}{12} \ln^2 x^2 M^2 + \frac{19}{36} \ln x^2 M^2}{x^2} + \right. \nonumber \\
&+& \left. \d_{\m \n} \Box \Box \frac{ - \frac{5}{12} \ln^2 x^2 M^2 - \frac{7}{36} \ln x^2 M^2}{x^2} \right] + \ldots
\er

\subsection{Diagram (h)}
\br
< B_{\m}^a(x) B_{\n}^b(y) >_{h} = &- g^4 & \int d^4 u  \; \D^{(c)}_{xu} \left[ f^{acx} f^{xdf} ( \d_{\m \s} \d_{\rho \eps} - \d_{\m \eps} \d_{\rho \s} ) + f^{adx} f^{xfc}  \times \right. \nonumber \\ 
&\times& \left. ( \d_{\m \eps} \d_{\s \rho} - \d_{\m \rho} \d_{\eps \s} ) + f^{afx} f^{xcd} ( \d_{\m \rho} \d_{\s \eps} - \d_{\m \s} \d_{\rho \eps} ) \right]  \times \nonumber \\ 
&\times& \D^{(d)}_{xu} f^{edc} \left[ \d_{\l \s} ( \stackrel{e}{\pa_{\rho}^{u}}- \stackrel{d}{\pa_{\rho}^u} )+ \d_{\l \rho} ( \stackrel{c}{\pa_{\s}^u} - \stackrel{e}{\pa_{\s}^u} ) + \d_{\s \rho} ( \stackrel{d}{\pa_{\l}^u} - \stackrel{c}{\pa_{\l}^u} ) \right] \times \nonumber \\ 
&\times& \D_{uy}^{(e)} f^{bfe} \left[ - 2 \d_{\n \l} D_{\eps}^y + \d_{\eps \l} ( \pa_{\n}^y - \stackrel{\leftarrow}{\pa_{\n}^y}) + 2 \d_{\n \eps} D_{\l}^y \right] \D_{xy} \nonumber 
\er 

In this expression we use the convention of  $\D^{(i)} \stackrel{i}{\pa_{\m}} \D^{(j)} = ( \pa_{\m} \D^{(i)} ) \D^{(j)}$. Evaluating all the index contractions
\br
<B_{\m}^a(x) B_{\n}^b (0) >_{h \; R} &=& - g^4 C_A^2 \d^{ab} \left[ \frac{45}{2} \pa_{\m} ( \D \pa_{\n} I^1 ) - 27 \D \pa_{\m} \pa_{\n} I^1 - 9 \d_{\m \n} \pa_{\l} ( \D \pa_{\l} I^1 ) \right]_R \nonumber \\
&=& \frac{g^4 C_A^2 \d^{a b}}{32 (4 \pi^2)^3} \left[ \pa_{\m} \pa_{\n} \Box \frac{ \frac{9}{4} \ln^2 x^2 M^2 + \frac{21}{4} \ln x^2 M^2}{x^2} + \right. \nonumber \\
&+& \left. \d_{\m \n} \Box \Box \frac{ - \frac{9}{4} \ln^2 x^2 M^2 + \frac{15}{4} \ln x^2 M^2}{x^2} \right] + \ldots
\er

\subsection{Diagram (i)}

This diagram and the next one have overlapping divergences as discussed with diagram (k). We will proceed as in that diagram, making use of the overlapping integrals listed in section \ref{Overlapping_div}.

The bare expression for this diagram is
\br
< B_{\m}^a (x) B_{\n}^b (y) >_i &=& - g^4 f^{afc} f^{gcd} f^{bde} f^{gef} \int d^4 u d^4 v \; \D_{xu} ( \stackrel{\leftarrow}{\pa_{\m}^x} - \pa_{\m}^x ) ( \pa_{\l}^u \D_{uy}) \times \nonumber \\
&\times& ( \pa_{\n}^y - \stackrel{\leftarrow}{\pa_{\n}^y}) \D_{yv} ( \pa_{\l}^v \D_{vx} ) \D_{uv}
\er

and operating this becomes
\br
< B_{\m}^a (x) B_{\n}^b (y) >_i = - \frac{1}{2} g^4 C_A^2 \d^{ab} &\left[ \right.& \pa_{\m}^x \pa_{\n}^y H[1, \pa_{\l} \; ; \; \pa_{\l},1] - 2 \pa_{\m}^x H[1, \pa_{\l} \; ; \; \pa_{\l} \pa_{\n} , 1] - \nonumber \\
& & \left. - 2 \pa_{\n}^y H[1, \pa_{\l} \pa_{\m} \; ; \; \pa_{\l},1] + 4 H[ 1, \pa_{\m} \pa_{\l} \; ; \; \pa_{\n} \pa_{\l} , 1] \; \right]
\er

Finally, with (\ref{int3}), (\ref{int6}), and (\ref{int15}) the renormalized expression is
\br
< B_{\m}^a (x) B_{\n}^b (0) >_{i \; R} &=& \frac{g^4 C_A^2 \d^{a b}}{32 (4 \pi^2)^3} \left[ \pa_{\m} \pa_{\n} \Box \frac{ - \frac{1}{12} \ln^2 x^2 M^2 - \frac{17}{36} \ln x^2 M^2 }{x^2} \right. + \nonumber \\
&+& \left. \d_{\m \n} \Box \Box \frac{ \frac{1}{12} \ln^2 x^2 M^2 + \frac{29}{36} \ln x^2 M^2}{x^2} \right] + \ldots
\er

\subsection{Diagram (j)}

This diagram is of the following form
\br
< B_{\m}^a (x) B_{\n}^b (y) >_j &=& 2 g^4 f^{acf} f^{cgd} f^{bde} f^{feg} \int d^4 u d^4 v \; \D_{xu} \left[ - 2 \d_{\m \s} D_{\rho}^x + \d_{\rho \s} ( \stackrel{\leftarrow}{\pa_{\m}^x} - \pa_{\m}^x ) + \right. \nonumber \\
&+& \left. 2 \d_{\m \rho} D_{\s}^x \right] ( \pa_{\rho}^u \D_{uy} ) ( \pa_{\n}^y - \stackrel{\leftarrow}{\pa_{\n}^y} ) \D_{yv} ( \pa_{\s}^v \D_{uv}) \D_{vx}
\er

Evaluating the index contractions this becomes
\br
< B_{\m}^a (x) B_{\n}^b (y) >_j = - g^4 C_A^2 \d^{ab} &\left[ \right.& - 4 \pa_{\n}^x \pa_{\l}^x H[1, \pa_{\m} \pa_{\l} \; ; \; 1 , 1] - 4 \pa_{\n}^x \pa_{\l}^x H[ 1, \pa_{\l} \; ; \; 1, \pa_{\m} ] - \nonumber \\
& & - 4 \pa_{\l}^x H[ 1, \pa_{\m} \pa_{\l} \; ; \; 1, \pa_{\n} ] - 4 \pa_{\l}^x H[ 1, \pa_{\l} \; ; \; 1, \pa_{\m} \pa_{\n} ] - \nonumber \\
& & - 4 \pa_{\n}^x H[ 1, \pa_{\m} \pa_{\l} \; ; \; 1, \pa_{\l} ] + \pa_{\m}^x \pa_{\n}^x H[ 1, \pa_{\l} \; ; \; 1, \pa_{\l}] - \nonumber \\
& & - 4 H[ 1, \pa_{\m} \pa_{\l} \; ; \; 1, \pa_{\n} \pa_{\l} ] + 4 \Box H[ 1, \pa_{\m} \; ; \; 1, \pa_{\n} ] + \nonumber \\
& & + 4 \Box H[ 1, 1 \; ; \; 1, \pa_{\m} \pa_{\n} ] - 2 \pa_{\n}^x H[ 1, \pa_{\m} \; ; \; 1, \Box] + \nonumber \\
& & + \pa_{\m}^x \pa_{\n}^x H[1, 1 \; ; \; 1, \Box] - 4 H[1, \pa_{\n} \Box \; ; \; 1, \pa_{\m} ] - \nonumber \\
& & \left. - 2 \pa_{\m}^x H[ 1, \pa_{\n} \Box \; ; \; 1, 1] \; \right]
\er

Proceeding in the same way as in diagram (k) we find the renormalized expression to be
\br
< B_{\m}^a (x) B_{\n}^b (y) >_j &=& \frac{g^4 C_A^2 \d^{a b}}{32(4 \pi^2)^3} \left[ \pa_{\m} \pa_{\n} \Box \frac{ \frac{1}{2} \ln^2 x^2 M^2 + \frac{1}{2} \ln x^2 M^2}{x^2} + \right. \nonumber \\
&+& \left. \d_{\m \n} \Box \Box \frac{ - \frac{1}{2} \ln^2 x^2 M^2 - \frac{1}{2} \ln x^2 M^2}{x^2} \right] + \ldots
\er

\section{Integral relations}
\label{ap_int_rel}

To deal with overlapping divergences, the following exact integral relations are very useful
\br
\int & d^4 u d^4 v &  \D_{xu} ( \pa_{\m}^x \D_{xv} ) ( \pa_{\l}^y \pa_{\n}^y \D_{yu} ) \D_{yv} \D_{uv} \label{rel_int1}\nonumber \\
&=& \pa_{\l}^x \int d^4 u d^4 v \; \D_{xu} ( \pa_{\m}^x \D_{xv} ) \D_{yu} ( \pa_{\n}^y \D_{yv} ) \D_{uv} + \nonumber \\
& & + \int d^4 u d^4 v \; \D_{xu} ( \pa_{\m}^x \D_{xv} ) \D_{yu} ( \pa_{\n}^y \pa_{\l}^y \D_{yv} ) \D_{uv} + \nonumber \\
& & + \pa_{\n}^x \int  d^4 u d^4 v \; \D_{xu} ( \pa_{\m}^x \D_{xv}) \D_{yu} ( \pa_{\l}^y \D_{yv} ) \D_{uv} + \nonumber \\
& & + \pa_{\n}^x \pa_{\l}^x \int d^4 u d^4 v \; \D_{xu} ( \pa_{\m}^x \D_{xv}) \D_{yu} \D_{yv} \D_{uv}  \nonumber \\
\\
\pa_{\l}^y \int & d^4 u d^4 v & \D_{xu} \D_{xv} ( \pa_{\l}^y \pa_{\n}^y \D_{yu} ) \D_{yv} \D_{uv} \label{rel_int2}\nonumber \\
& = & \frac{1}{2} \pa_{\n}^y \int d^4 u d^4 v \; \D_{xu} \D_{xv} ( \Box \D_{yu} ) \D_{yv} \D_{uv} - \nonumber \\
& & - \int d^4 u d^4 v \; \D_{xu} \D_{xv} ( \Box \D_{yu} ) \D_{yu} \D_{uv} + \nonumber \\
& & + \frac{1}{2} \pa_{\n}^y \pa_{\l}^y \int d^4 u d^4 v \; \D_{xu} \D_{xv} ( \pa_{\l}^y \D_{yu} ) \D_{yv} \D_{uv} \nonumber \\
\\
\pa_{\l}^y \int & d^4 u d^4 v& \D_{xu} \D_{xv} ( \pa_{\l}^y \pa_{\n}^y \D_{uy} ) ( \pa_{\m}^y \D_{yv} )\D_{uv} \label{rel_int3} \nonumber \\
&=& \frac{1}{4} \pa_{\n}^y \pa_{\m}^y \pa_{\l}^y \int d^4 u d^4 v \; \D_{xu} \D_{xv} ( \pa_{\l}^y \D_{uy} ) \D_{yv} \D_{uv} + \nonumber \\
& & + \frac{1}{4} \pa_{\n}^y \Box \int d^4 u d^4 v \; \D_{xu} \D_{xv} ( \pa_{\m}^y \D_{uy} ) \D_{yv} \D_{uv} - \nonumber \\
& & - \frac{1}{2} \Box \int d^4 u d^4 v \; \D_{xu} \D_{xv} ( \pa_{\m}^y \pa_{\n}^y \D_{yu} ) \D_{yv} \D_{uv} \nonumber \\
\er

To prove these relations, we have only to perform the derivatives and expand the different terms.


\end{document}